\documentclass{elsart}
\input{epsf}
\newcommand{\lpat}{\ell_{\rm pattern}}
\newcommand{\tfront}{\tau_{\rm  front}}

\begin{document}
\begin{frontmatter}
\title{Three basic issues concerning interface dynamics in nonequilibrium
  pattern formation }
\author{ Wim van Saarloos}
\address{Instituut-Lorentz, Leiden University, P.O. Box 9506, 2300 RA
  Leiden,\\ The Netherlands}

\begin{abstract}
In these lecture notes, we discuss at an elementary level three themes
concerning interface 
dynamics that play a role in pattern forming systems: {\em (i)} We
briefly review three examples of systems in which the normal growth velocity
is proportional to the gradient of a bulk field which itself obeys a
Laplace or diffusion type of equation (solidification, viscous fingers
and streamers), and then discuss why the Mullins-Sekerka instability
is common to all such gradient systems. {\em (ii)} Secondly, we
discuss how underlying an effective interface description of systems
with smooth fronts or transition zones, is the assumption that the
relaxation time of the appropriate order parameter field(s) in the
front region is much smaller than the time scale of the evolution of
interfacial patterns. Using standard arguments we illustrate that this
is generally so for fronts that separate two (meta)stable phases: in
such cases, the relaxation is typically exponential, and the  relaxation
time in the usual models goes to zero in the limit in which the front
width vanishes. {\em (iii)} We finally summarize recent results that
show that so-called ``pulled'' or ``linear marginal stability'' fronts
which propagate into unstable states have a  very slow  universal
power law relaxation. This slow relaxation makes the usual ``moving boundary'' or ``effective
interface'' approximation for problems with thin fronts, like streamers,
impossible.
\end{abstract}
\end{frontmatter}

\section*{Introduction}
In this course, and in the two related lectures by Ebert and Brener on
their work in \cite{streamers} and \cite{brener}, some basic features
of the dynamics of growing interfaces in systems which spontaneously
form nonequilibrium patterns will be discussed.  The analysis of such
growth patterns has been an active field of research in the last
decade. Moreover, the field is quite diverse, with examples coming
from various (sub)disciplines within physics, materials science, and
even biology --- combustion, convection, crystal growth, chemical
waves in excitable media and 
the formation of Turing patterns, dielectric breakdown, fracture,
morphogenesis, etc.  We therefore can not hope to
review the whole field, but instead will content ourselves with
addressing three rather basic topics which we consider to be of rather
broad interest, in that they appear (in disguise) in many areas of
physics and in some of the related fields. These three themes are
explained below. 

Our {\em first theme} concerns the {\em generality of interfacial
  growth problems in which the normal growth velocity $v_n$ is
  proportional to the gradient $\nabla \Phi$ of a bulk field $\Phi$
  ($v_n \sim \nabla \Phi$), and the associated long wavelength
  instability of such interfaces.} As we shall see, one important
class of interfacial growth problems with these properties is
diffusion limited growth: either the interface grows through the
accretion of material via diffusion through one of the adjacent bulk
phases, or the growth of the interface is limited by the speed through
which, e.g., heat can be transported away from the interface through
diffusion. Since the diffusion current near the interface is
proportional to the gradient of the appropriate field in the bulk (a
concentration field, the temperature, etc.), $v_n \sim \nabla \Phi$ in
such growth problems. But these are not the only possibilities of how
one can have an interface velocity proportional to $\nabla \Phi$. As we
will discuss, in viscous fingering an air bubble displaces fluid
between two closely spaced plates; as the fluid velocity between the
plates is proportional to the gradient of a pressure field $p$, the
fact that the air displaces all the fluid means that again at the
interface $v_n \sim \nabla p$. Likewise, for an ionization front the
interface velocity is determined mostly by the drift velocity of
electrons in the electric field ${\rm E} = -\nabla \Phi$, where now
$\Phi $ is the electrical potential, and so again $v_n \sim |\nabla
\Phi|$.  As we shall see, all such interfacial growth problems where
the bulk field itself obeys a Laplace of diffusion equation, exhibit a
long wavelength instability, the so-called Mullins-Sekerka instability
\cite{mullins}.  This instability lies at the origin of the formation
of many nontrivial growth patterns, and Brener and Ebert discussed
examples of these at the school.  Although this theme is not at all
new, it is nevertheless useful to discuss it as an introduction and to
stress the generality --- the same Mullins-Sekerka instability plays a
role in fractal growth processes like Diffusion Limited Aggregation
\cite{dla}. The above issues are the subject of the first lecture and
section \ref{theme1}.

In physics, it is quite common --- and often done intuitively without
even stating this explicitly --- to switch back and forth between a
formulation in terms of a mathematically sharp interface (an
infinitely thin surface or line at wich the physical fields or their
derivatives can show discontinuities) and a formulation in terms of a
continuous order parameter field which exhibits a smooth but
relatively thin transition zone or domain wall. E.g., we think of the
interface between a solid and its melt as a microscopically thin
interface, whose width is of the order of a few atomic dimensions.
Accordingly, the formulation of the equations that govern the
formation of growth patterns of a solid which grows into an
undercooled melt on much larger scales $\lpat$ have traditionally been
formulated in terms of a sharp interface or boundary. The equations,
which will be discussed below, are then the diffusion equation for the
temperature in the bulk of the liquid and the solid, together with
boundary conditions at the interface. These interfacial boundary
conditions are a kinematic equation for the growth velocity of
the interface in terms of the local interface temperature, and a
conservation equation for the heat. The latter expresses that the
latent heat released at the interface upon growth of the solid has to
be transported away through diffusion into the liquid and the solid.
In other words, in an interfacial formulation, the appropriate
equations for the bulk fields are introduced, but the way in which the
order parameter changes from one state to the other in the interfacial
region, is not taken into account explicitly\footnote{Note that if we
  consider a solid-liquid interface of a simple material so that the
  interface width is of atomic dimensions, there can be microscopic
  aspects of the interface physics that have to be put in by hand in
  the interfacial boundary conditions anyway, as they can not really
  be treated properly in a 
  continuum formulation. E.g., if the solid-liquid interface is rough,
  a linear kinetic law in which the interface grows in proportion to
  the local undercooling is appropriate. If the interface is faceted,
  however, a different boundary condition will have to be used.}: the
physics at the interface is lumped into appropriate boundary
conditions. Such an interface formulation of the equations often
expresses the physics quite well and most efficiently, and is often
the most convenient one for the analytical calculations which we will
present later. For numerical calculations, however, the existence of a
sharp boundary or interface is a nuisance, as they force one to
introduce highly non-trivial interface tracking methods.  Partly to
avoid this complication, several workers have introduced in the last
few years different models, often referred to as phase-field
models \cite{fife,karma}, in 
which the transition from one phase or state to another one is
described by introducing a continuum equation for the appropriate
order parameter. Instead of a sharp
interface, one then has a smooth but thin transition zone of width
$W$, where the order parameter changes from one (meta)stable state to
another one. Numerically, such smooth interface models are much easier
to handle, since one can in principle apply standard numerical
integration routines\footnote{While the numerical code may be
  conceptually much more straightforward, the bottleneck with these
  methods is that one now needs to have a small gridsize,
  so as to properly resolve the variations of the order parameter
  field on the scale $W$. At the same time, one usually wants to study
  pattern formation on a scale $\lpat \gg W$, so that many gridpoints
  are needed. Hence computer power becomes the limiting
  factor. Nevertheless, numerical similations of dendrites using such
  phase-field models nowadays appear to present the best way to test
  analytical predictions and to compare with experimental data \cite{karma}. }.

While we mentioned above an example where one has (sometimes in a
somewhat artifial way) introduced continuum field equations to analyze
a sharp interface problem, insight into the dynamics of problems with
a moving smooth but thin transition zone is often more easily gained
by going in the opposite direction, i.e., by viewing this zone as a
mathematically sharp interface or shock front. An example of this is
found in combustion \cite{comb1,comb2}. In premixed flames, the
reaction zone is usually quite small, and one speaks of flame sheets.
Already long ago, Landau (and independently Darrieus) considered the
stability of planar flame fronts by viewing them as a sharp
interface \cite{comb1,comb2}. In the last 20 years, much progress has
been made in the field of combustion by building on this idea of using
an effective interface description of thin flame sheets. Likewise, much of the
progress on understanding chemical waves, spirals and other patterns
in reaction diffusion systems rests on the
possibility to exploit  similar ideas \cite{meron,goldstein2}. Other
examples from 
condensed matter physics: if the magnetic anisotropy is not too large,
domain walls in solids can have an appreciable width, but for many
studies of magnetic domains of size much larger than this width, we
normally prefer to think of the walls as being infinitely
thin \cite{kleman}. Similar considerations hold for domains in
liquid crystals \cite{kleman}. In studies of coarsening (the gradual
increase in the typical length scale after a quench of a binary fluid
or alloy into the so-called spinodal regime where demixing occurs),
both smooth and sharp interface formulations are being
used \cite{gunton,bray,langer2,allen}.

At a summer school on statistical physics, it seems appropriate to
note in passing that some of the model equations which include the
order parameter are very similar to those studied in particular in the
field of dynamic critical phenomena, such as model {\em A, B,}.. etc.
in the classification of Hohenberg and Halperin \cite{hohandhal}.
Here, however, we are not interested in the universal scaling
properties of an essentially homogeneous system near the critical
point of a second order phase transition, but in many cases in the
nonlinear nonequilibrium dynamics of interfaces between a metastable
and a stable state. This corresponds to the situation near a first
order transition\footnote{ This will be illustrated, e.g., by the
  Landau free energy $f$ in Fig.\ 4({\em b}) below. Here the states
  $\phi=0$ and $\phi=\phi_s$ both correspond to minima of the free
  energy density $f$. In section 2.1 fronts between these two
  (meta)stable states are discussed.}. When we consider in section
\ref{theme3} fronts which propagate into an unstable state, this can
be viewed as the interfacial analogue of the behavior when we quench a
system through a second order phase transition, especially within a
mean-field picture\footnote{In the mean field picture, the Landau free
  energy has one (unstable) maximum at $\phi=0$ and one minimum at
  some $\phi=\phi_s \neq 0$. Fronts propagating into
  an unstable state precisely correspond to fronts between these two
  states. Compare Fig.\ 9({\em b}), where $V=-f$.}, where fluctuations
are not important\footnote{It is actually rather exceptional to have
  propagating interfaces when we quench a system through the
  transition temperature of a second order phase transition, because
  the fluctuations make it normally impossible to keep the system in
  the phase which has become unstable long enough that propagating
  interfaces can develop. Nevertheless, there is one example of a
  thermodynamic system in which the properties of propagating
  interfaces were used to probe the order of the phase transition: for
  the nematic--smectic-A transition, which was predicted to be always
  weakly first order, the dynamics of moving interfaces was used to
  probe experimentally \cite{anisimov} the order of the transition
  close to the point which earlier had been associated with a
  tricritical point (the point where a second order transition becomes
  a first order transition). These dynamical interface measurements
  confirmed that the transition was always weakly first order
  \cite{anisimov}. Note finally that in pattern forming systems, the
  fluctuations are often small enough (see the remarks about this in
  the next paragraph of the main text) that fronts propagating into
  unstable states can be prepared more easily. For an example of such
  fronts in the Rayleigh-Benard instability, see \cite{fineberg}.}.

Finally, we also stress that while thermal fluctuations are essential
to second order phase transition, they can often be neglected in
pattern forming systems, since the typical length and energy scales of
interest in pattern forming systems are normally very large (See
section VI.D of  \cite{crosshoh} for further discussion of this point).

A word about nomenclature. For many physicists, front, domain wall and
reaction zone are words that have the connotation of describing smooth
transition zones of finite thickness\footnote{And even this is not
  true: in adsorbed monolayers walls usually have only a microscopic
  thickness; e.g., light and heavy walls are concepts that
  have been introduced to distinguish walls which differ in the atomic
  packing in one row. See, e.g.,  \cite{dennijs}.}, while the word
interface is being 
used for a surface whose thickness is so small that it can be treated
as a mathematically sharp boundary of zero thickness. The
approximation in which the interface thickness is taken to zero is
sometimes refered to as a {\em moving boundary approximation}. Since
neither this concept nor the meaning of the word ``interface'' is
universally accepted, we will 
sometimes use {\em effective interface description} or {\em effective
  interface approximation} as an alternative to ``moving boundary
approximation'' to denote a description of a front or
transition zone by a sharp interface with appropriate boundary
conditions. 

Of course, switching back and forth between a sharp interface
formulation or one with a smooth continuous order parameter field is
only possible if the latter reduces to the first in the limit in which
the interface thickness
$W\rightarrow 0$ (the interface width is illustrated in Fig.\ 8
below). Indeed, it is possible to derive an effective 
interface description systematically by performing an expansion of the
equations in powers of $W$ (technically, this is done using singular
perturbation theory or matched asymptotic
expansions \cite{fife,karma,bender,vandyke,fife2}). In such an
analysis, the wall or front is treated as a sharp interface when
viewed on the ``outer'' pattern forming length scale $\lpat \gg W$,
and the dynamics of the front or wall on the ``inner'' scale $W$
emerges in the form of one or more boundary conditions for this
interface. E.g., one boundary condition can be a simple expression for
the normal velocity of the interface in terms of the local values of
 the slowly varying outer fields, like the temperature.

This brings us to the {\em second theme} and lecture of this course:
{\em for the adiabatic decoupling of slow and fast variables
  underlying an effective interface description, an exponential
  relaxation of the front structure and velocity is necessary}. The
point is that an interfacial description --- mapping a smooth
continuum model onto one with a sharp interface for the analysis of
patterns on a length scale $\lpat \gg W$ --- is only possible if we
can make an {\em adiabatic decoupling}. In intuitive terms, this means
that if we ``freeze'' the slowly varying outer fields (temperature,
pressure, etc.) at their instantaneous values and perturb the front
profile on the scale $W$ by some amount, then the front profile and
speed should relax as $\exp(-t/ \tfront)$ to some asymptotic shape and
value which are given in terms of the ``frozen'' outer field. If,
moreover, the inner front relaxation time $\tfront$ vanishes as
$W\rightarrow 0$ (e.g., in the model we will discuss $\tfront \sim W
$), then indeed in the limit $W \rightarrow 0$ the relaxation of the
order parameter dynamics within the front region decouples completely
from the slow time and length scale variation of the outer fields, as
in the limit $W\rightarrow 0$ both the length and the time scale
become more and more separated. The adiabatic decoupling then implies
that for $W \ll \lpat$ the front follows essentially instantaneously
the slow variation of the outer fields in the region near the front.
Accordingly, in the interfacial limit $W\rightarrow 0$ the front
dynamics on the inner scale $W$ then translates into boundary
conditions that are {\em local in time and space} at the interface. As
we shall illustrate, for fronts between two (meta)stable states, the
separation of both length and time scale as $W\rightarrow 0$ is
normally the case, and this justifies an interfacial description.

Of course, stated this way, the above point may strike you as
trivial, as it is a common feature of problems in which fast variables
can be eliminated \cite{kampen}. However, it is an observation that we
have hardly ever seen stressed or even discussed at all in the
literature, and its importance is illustrated by our {\em third theme:
  fronts propagating into an unstable state may show a separation of
  spatial scales in the limit $W\rightarrow 0$, but need not show a
  separation of time scales in this limit.} Our reason for the last
statement is that, as we will discuss, a wide class of fronts which
propagate into an unstable state (the interfacial analogue of the
situation near a second order phase transition) exhibits slow power
law relaxation ($\sim 1/t$). This certainly calls the possibility of
an effective interface formulation with boundary conditions which are
local in space {\em and} time into question, but the consequences of
this power law relaxation still remain to be fully explored.

The connection between the issue of the front relaxation and the issue
of the separation of time scales necessary for an effective interface
description is still a subject of ongoing research of Ebert and
myself. We will in these lecture notes only give an introduction to
the background of this issue and to the ideas underlying the usual
approaches, leaving the real analysis and a full discussion of this
problem to our future publications \cite{ebert2}.

That our third theme is not a formal esotheric issue, is illustrated
by the fact that it grew out of our attempt to develop an interfacial
description for streamers. As has been discussed by Ebert in her
seminar, streamers are examples of a nonequilibrium pattern forming
phenomenon. They consist of a very sharp fronts ($W \approx 10 \mu m$)
which shows patterns with a size $\lpat$ of order 1 mm \cite{vitello}.
However, a streamer front turns out to be an example of a front
propagating into an unstable state  \cite{streamers}, and we have found
through bitter experience that the standard methods to arrive at an
interfacial approximation break down, and that the slow power law
relaxation lies at the heart of this. Apart from this, the power law
relaxation is of interest in its own right, especially at a summer
school on Fundamental Problems in Statistical Mechanics, as its
universality is  reminiscent of the universal behavior near
a second order critical point in the theory of phase transitions ---
the common origin of both is in fact the universality of the flow near
the asymptotic fixed point.

\section{Gradient driven growth problems and the Mullins-Sekerka
  instability}\label{theme1}
\subsection{The dendritic growth equations  \cite{langer,caroli,kassner}}
\begin{figure}
\epsfysize=5.5cm
\hspace*{10.2cm}
\epsffile{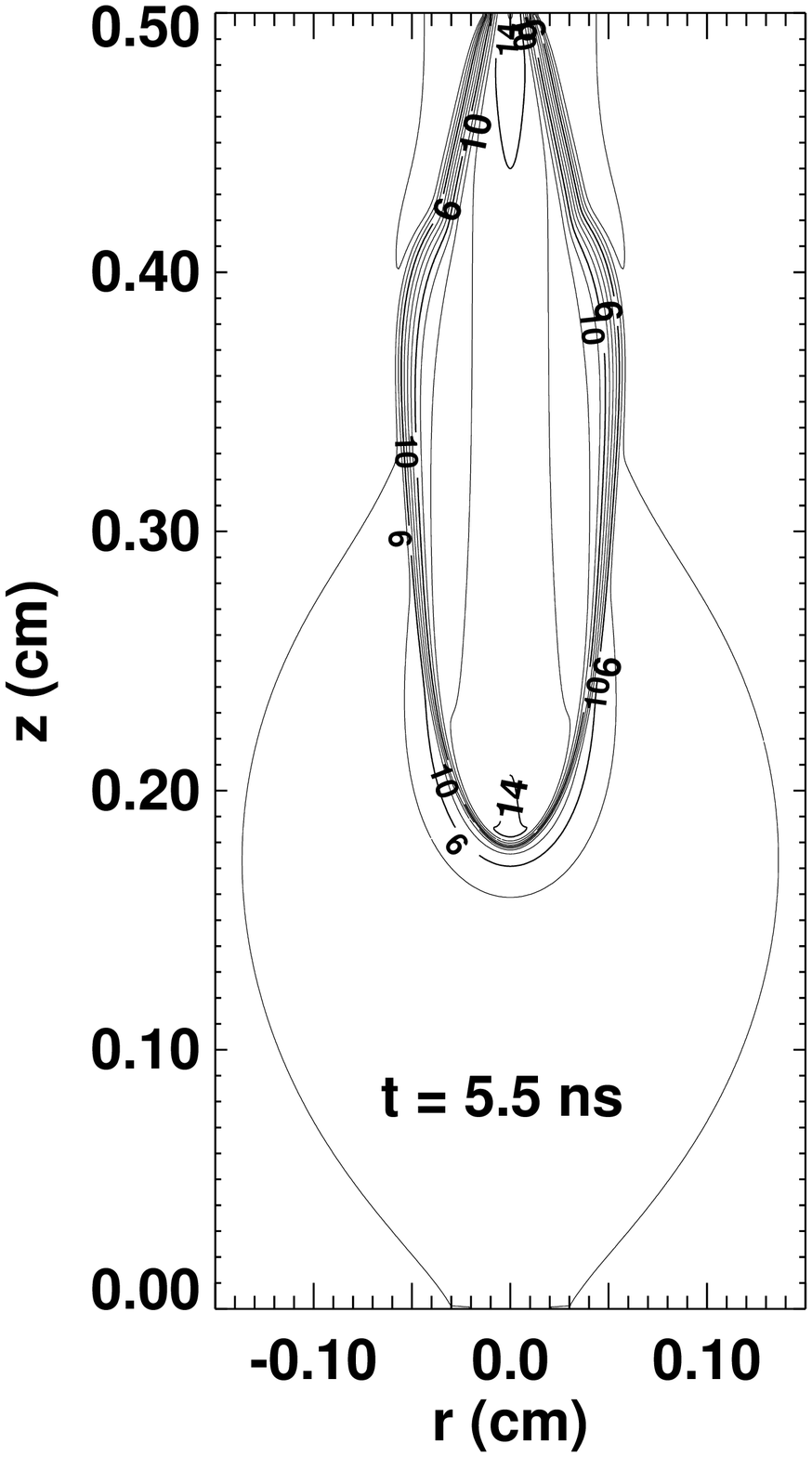}
\caption{\footnotesize\label{dendretc}Examples of three growth patterns: a dendrite
  in {\em (a)}, a viscous finger in {\em (b)}, and streamer in {\em
    (c)} . Usually, dendritic growth is studied in liquids with a
  melting temperature near room temperature. The one shown in {\em
    (a)} was observed in $^3He$ \cite{rolley} at 100 mK, and the fact
  that it is similar in shape and form to those usually observed
  illustrates the generality of dendritic growth [courtesy of
  E. Rolley and S. Balibar]. In {\em (b)}, a top
  view of a viscous finger is shown. The air inside the finger like
  pattern displaces the oil outside (from \cite{tabeling}, with
  permission from the author). The
  streamer pattern is from a numerical simulation \cite{vitello}. }
\end{figure}

When we undercool a pure liquid below the melting temperature, the
liquid will not solidify immediately. This is because below the
melting temperature the liquid is only metastable. Moreover, the
solid-liquid transition is usually strongly first order, so that the
nucleation rate for the solid phase to form through nucleation at
small to moderate undercoolings is low. If, however, we bring a solid
nucleus into the melt, the solid will start to grow immediately at its
interfaces. Initially the shape of the solid germ remains rather
smooth (we assume the interface to be rough, not faceted \cite{weeks}),
but once it has grown sufficiently large,
it does not stay rounded (like an ice cube melting in a soft-drink),
but instead branch-type structures grow out. An example of such a
so-called dendrite is shown in Fig. \ref{dendretc}{\em (a)}. The basic
instability underlying the formation of these dendrites is the
Mullins-Sekerka instability discussed below, and which in this case is
associated with the build-up near the interface of a diffusion
boundary layer in the temperature. This in turn is due to the fact
that while the solid grows, latent heat is released at the
interface. In fact, the amount of heat released is normally so large
that most of the heat has to diffuse away, in order to prevent the local
temperature to come above the melting temperature $T_M$. E.g., for
water the latent heat released when a certain volume solidifies is
enough to heat up that same volume by about 80 $^o$C. So, since the
undercooling is normally just a few degrees, most of the latent heat
has to diffuse away in order for the temperature not to exceed $T_M$.

The basic  equations that model this  physics  are the diffusion
equation for the temperature in the liquid and the solid,
\begin{equation} \label{dendrdiff}
{{\partial T}\over{\partial t}} = D \nabla^2 T ~,
\end{equation}
together with the boundary conditions at the interface
\begin{eqnarray}\label{dendrbc1}
\hspace*{2.8cm} {{L}\over{c}} v_n & = & -D [ (\nabla T^\ell)_{n,int} - (\nabla
T^s)_{n,int} ] ~,
\\
v_n & = &  {{1}\over{\beta}} [ T_M(1-(\sigma/L) \kappa) - T^{\rm
  int}]~. \label{dendrbc2}
\end{eqnarray}

Eq. (\ref{dendrdiff}) is just the normal heat diffusion equation for
the temperature; it holds both in the liquid ($T=T^\ell$) and in the
solid ($T=T^s$). At the interface, the temperature is continuous, so
 there $T^\ell=T^s=T^{\rm int}$. In (\ref{dendrdiff}), we have for
simplicity taken the diffusion coefficient $D$ in both phases
equal. The first boundary condition (\ref{dendrbc1}) expresses the
heat conservation  at the interface: $v_n$ is the normal growth
velocity of the interface, so $Lv_n$ is the amount of heat released
at the surface per unit time ($L$ is the latent heat per unit volume).
If we consider an infinitesimal ``pillbox'' at the interface, the heat
produced has to be equal to the net amount of heat which is being
transported out of the flat sides through heat diffusion. The heat
current is in general $-cD \nabla T$, with $c$ the specific heat per
unit volume (the combination $cD$ is the so-called  heat
conductivity),  and we denote the components normal to the interface
by $(\nabla T^\ell)_{n,int}$ and $ (\nabla T^s)_{n,int}$. After
dividing by $c$, we therefore recognize in the right hand side of
(\ref{dendrbc1})  the net heat flow away from the interface. Note that
this equation is completely fixed by a conservation law at the
interface, in this case conservation of heat, and that it can be
written down by inspection. The only input is the assumption that the
interface is very thin. Moreover, it shows the structure we mentioned
in the beginning, namely that the normal growth velocity is
proportional to the gradient of a bulk quantity --- the larger the
difference in the gradients in the solid and the liquids is, the
faster the growth can be. 

Finally, the second boundary condition at the liquid-solid interface
(\ref{dendrbc2}) is essentially the local kinetic equation which
expresses the microscopic physics at the interface: We 
assume the interface to be rough, so that the interface can be
smoothly curved  \cite{weeks}. $T_M[1-(\sigma/L) \kappa]$ is then the melting
temperature of such a  curved interface, where $\sigma $ is the
surface tension. Here the  curvature $\kappa$
is taken positive when the solid bulges out into the liquid; the
suppression of $T_M$ in such a case can intuitively be thought of as
being due to the fact that there are more broken ``crystalline'' bonds if the solid is
curved out into the liquid, but the relation follows quite generally
from thermodynamic considerations  \cite{langer}. The ratio $(\sigma/L)$
 is a small microscopic length, say of the order of
tens of {\AA}ngstroms.  The necessity of
introducing the suppression of the local interface melting temperature
will emerge later, when we will see that if we don't do so, there
would be a strong short wavelength (``ultraviolet'') instability at
the interface\footnote{Due to the crystalline anisotropy, the
  capillary parameter actually depends on the angles the interface
  makes with the underlying crystalline lattice. It has been
  discovered theoretically that this crystalline anisotropy actually
  has a crucial influence on dendritic growth: without this
  anisotropy, needle-like tip solutions of a dendrite don't exist, and
  the growth velocity of such needles is found to scale  with a 7/4 power of the
  anisotropy amplitude. We refer to  \cite{langer,kassner,reviews} for
  a detailed discussion of
  this point.}.

Now that we understand the meaning of $T_M(1-(\sigma/L) \kappa)$ as a local
melting term of a curved interface, we see that Eq. (\ref{dendrbc2})
just expresses the linear growth law for rough interfaces
 \cite{weeks}; $1/\beta$ in this expression has the meaning of a
mobility. If we take the limit of infinite mobility ($1/\beta
\rightarrow \infty$), the interface grows so easily that we can
approximate (\ref{dendrbc2}) by
\begin{equation}\label{localeq}
\hspace*{-1.5cm} (\beta \rightarrow 0) \hspace*{1.5cm} T^{\rm int} =
T_M [1 -(\sigma/L)  \kappa]~,
\end{equation}
which is sometimes refered to as the local equilibrium approximation.

We stress that the boundary condition (\ref{dendrbc2}) is {\em local
  in space and time}, i.e., the growth velocity $v_n$ responds {\em
  instantaneously} to the local temperature and curvature. There are of
course sound physical reasons why this is a good approximation:
the typical solid-liquid interfaces we are interested in are just a
few atomic dimensions wide, and respond on the time-scale of a few
atomic collision times (of order picoseconds) to changes in
temperature \cite{gilmer}, while Eqs. (\ref{dendrdiff})-(\ref{dendrbc2})
are used to analyze pattern formation on length scales of order
microns or more and with growth velocities of the order of a $\mu$m/s,
say. Hence, an interface grows over a distance comparable to its width
in a time of the order of $10^{-3}$ seconds, and the time scale for
the evolution of the patterns is typically even slower.  This
wide separation of length and time  scales justifies  the assumptions
underlying the interfacial boundary conditions.

Eqs. (\ref{dendrdiff})-(\ref{dendrbc2}), together with appropriate
boundary conditions for the temperature far away from the interface,
constitute the basic equations that describe the growth of a dendrite
into a pure melt. They may look innocuous, as they appear to be linear
equations, but they are not! The reason is that they involve the
unknown position and shape of the interface through the boundary
conditions, and that the dynamics of the interface depends in turn on
the diffusion fields: the
location of the interface has to be found self-consistently in the
course of solving 
these equations! This is why such a so-called {\em moving boundary
  problem} 
is so highly nonlinear and complicated.

These equations (with crystalline anisotropy included
in the capillary term $(\sigma/L)$) were actually the starting point of
Brener's talk at the school, and the work he  discussed  \cite{brener}  showed
how challenging 
the nonlinear analysis necessary to obtain a phase diagram of growth
patterns can be. Such work builds on many advances made in the last
decade on understanding the growth velocity and shape of
dendrites, and for a discussion of these we refer to the
literature \cite{reviews,brener2}.

\subsection{Viscous fingering \cite{reviews}}
In viscous fingering, or Saffman-Taylor fingering, one considers a
fluid (typically an oil) confined between two long parallel closely
spaced plates. In Fig.\ 1({\em b}) one looks from the top at such a
cell --- the plates, separated by a small distance $b=0.8 ~mm$, are
thus in the plane of the paper. The two black 
sides in this photo constitute the lateral side walls; the oil between
the plates can not penetrate into these. The distance between these
side walls is 10 {\em cm} , hence the lateral width of the cell is much
smaller than the spacing $b$ of the plates. The thin line is the 
air-fluid interface and the region inside the finger-like shape is
air, while the oil is outside.  The air is blown into the area between
the plates from the upper part of the figure. If the
air-fluid interface initially stretches all the way across the cell
from left to right, one quickly finds that when the air
is blown in, this interface is unstable, and that after a while a
single finger-shaped pattern like the one shown in Fig.\ 1({\em b})
penetrates into the fluid. Understanding 
the shape and width of this finger has been a major theme in
interfacial pattern formation \cite{reviews}. In simple fluids it is
so well understood that the analysis of the finger shapes when
surfactants or polymers are added to the displaced fluid has become a
way to learn something about the resulting properties of the fluid and
the air-fluid interface \cite{bonn}. We will content ourselves here
with giving an introduction of the basic equations, aimed at bringing
out the same gradient-driven structure of the interface equations.

In viscous fingering experiments, the spacing $b$ between plates is
much smaller than the width of the cell (the distance between the two
dark sides in Fig.\ 1({\em b}). As a result, the average
fluid flow field varies in the plane of the cell only slowly over
distances of the order of the lateral dimensions of the cell. Locally,
therefore, the flow in the small direction normal to the planes is
almost like that of homogeneous planar Poisseuille flow, for which we
know that the average fluid velocity is $-b^2/(12 \eta)$ times the
gradient of the pressure $p$. Here $\eta$ is the kinematic viscosity
of the fluid. Hence, if we now introduce ${\bf v} (x,y)$ as the
height-averaged flow field between the plates of the cell, which we
take to lie in the $xy$-plane, we have ${\bf v} (x,y)=
-b^2/(12 \eta) \nabla p$. Taking the fluid to be incompressible,
$\nabla \cdot {\bf v}=0$, implies that in the bulk of the fluid the
pressure simply obeys the Laplace equation,
\begin{equation}\label{viscbulk}
\nabla^2 p =0~,
\end{equation}
while at the interface the fact that the air displaces all the fluid
is expressed by
\begin{equation}\label{viscbc1}
v_n = - {{b^2}\over{12 \eta}} (\nabla p )_{n,int}~.
\end{equation}
Furthermore, if we ignore the viscosity of the air and wetting effects, the pressure at
the interface is nothing but the equilibrium pressure  of a smoothly
curved air-fluid 
interface, i.e.,
\begin{equation} \label{viscbc2}
p |_{int} = p_0 -\sigma \kappa~,
\end{equation}
where as before $\sigma $ is the surface tension and $p_0$ the background
pressure in the gas. The curvature term $-\sigma \kappa$ is the direct
analogue of the one in (\ref{dendrbc2}); in the context of air-fluid
interfaces it is known as the Laplace pressure term, and it
corresponds to the well-known effect that the pressure inside a soap
bubble is larger than the one outside. As in the case of crystal
growth discussed above, the form of the boundary conditions can
essentially 
be guessed on physical grounds and by appealing to the fact that the
microscopic relaxation time at the interface is typically orders of
magnitude smaller than the time and length scale of the pattern.

\subsection{Streamer dynamics --- a moving charge sheet?}
We finally introduce a problem which is not at all understood in
detail but whose similarity with dendrites and viscous fingers
motivated some of the issues discussed here  \cite{streamers}. The
basic phenomenon is that when an electric field is large enough,  an
electron avalanche type can build up in a gas, due to the fact that
free electrons get accelerated  sufficiently that they ionize neutral
molecules, thus generating more free electrons, etc. Streamers are the
type of dielectric breakdown fronts that can occur in gases as a
combined 
result of this avalanche type of phenomenon and the screening of the
field due to the build-up of a charge layer. The basic
equations that are being used to model this behavior
 are the following continuum balance equations for the
electron  density $n_e$ and ion density $n_+$, 
  and the electric field ${\bf E}$  \cite{vitello}
\begin{eqnarray}
\label{stream1}
\hspace*{2.8cm} \partial_t\;n_e \;+\; \nabla \cdot{\bf j}_e
&=& |n_e \mu_e {\bf E}| \;\alpha_0
\;\mbox{e}^{-E_0/|{\bf E}|}~,
\\
\label{stream2}
\partial_t\;n_+ \;+\; \nabla \cdot{\bf j}_+
&=& |n_e \mu_e {\bf E}| \;\alpha_0
\;\mbox{e}^{-E_0/|{\bf E}|}~,
\end{eqnarray}
and the Poisson equation
\begin{equation}
\label{stream3}
\nabla \cdot {\bf E} = {{e}\over{\varepsilon_0}} \;(n_+ -n_e)
~.
\end{equation}
The electron and ion current densities ${\bf j}_e$ and ${\bf j}_+$ are
\begin{equation}
\label{stream4}
{\bf j}_e = - n_e \;\mu_e \;{\bf E} - D_e \;\nabla \;n_e ~,\hspace*{1cm}
{\bf j}_+ = 0~,
\end{equation}
so that ${\bf j}_e$ is the sum of a drift and a diffusion term, while
the ion current ${\bf j}_+$ is neglected, since the ions are much less
mobile than the electrons.  The right hand side  of Eqs.\ (\ref{stream1}) and
  (\ref{stream2}) is a 
source term due to the ionization reaction: In high fields free
electrons can generate free electrons and ions by impact on neutral
molecules. The source term is given by the magnitude of the electron
drift current times the target density times the effective ionization
cross section; the constant $E_0$ in the ionization probability
depends on the mean free path of the electrons and the ionization
energy of the neutral gas molecules. The rate constant $\alpha_0$ has
the dimension of an 
inverse length. The exponential function expresses, that only in
high fields electrons have a nonnegligible probability to collect
the ionization energy between collisions.

In Fig. \ref{dendretc}({\em c}), we show a plot of the simulations of
the above equations \cite{vitello}. In this figure, a gap between two
planar electrodes across which there is a large voltage difference is
studied for parameters in the model that correspond to $N_2$ gas.
Initially, the electron density is essentially zero everywhere except
in a very small region near the upper electron. The simulation of Fig.
\ref{dendretc}({\em c}) shows the situation 5.5 nanoseconds later; the
lines in this plot are lines of constant electron density. The density
differs by a factor 10 between successive lines. Since these lines are
closely spaced --- the electron density rises by a factor $10^{10}$ in
a few $\mu$m --- the simulations illustrate that a streamer consists
of an ionized region (inside the contour lines) propagating into a
non-ionized zone. Inside this zone, almost all of the ionization takes
place, and the total charge density is nonzero. It is this nonzero
charge density that also screens the electric field from the interior
of the streamer. We can therefore also think of the streamer as a
moving charge sheet, whose shape is somewhat like a viscous finger. In
fact, if the upper electrode is spherical rather than planar, one
finds branched streamers which are reminiscent of dendrites
\cite{vitello2}. Note that in this interfacial picture, the charge
sheet is also the reaction zone where most of the ionization takes
place, and that the build-up of the charge in this zone is at the same
time is responsible for the screening of the field in the interior of
the streamer.

An immediate question that comes to mind is whether we can analyze
such a streamer as a moving interface problem by mapping the continuum
equations onto a a sharp interface of zero thickness, by taking the
limit in which the charge sheet becomes infinitely thin
\cite{streamers}. If such an analysis can be done, very much along the
lines of the analysis for combustion or for the so-called phase-field
models mentioned in the introduction, one would intuitively expect
that the dynamics in the transition zone would, in the $W\rightarrow
0$ limit, translate into boundary conditions at the interface. In
particular, one expects one equation expressing charge conservation,
and a kinematic relation for the normal velocity of the interface.
Based on ones experience with the other problems described earlier,
this kinematic expression might be guessed to express the local
interface velocity as a function of the instantaneous values of the
``outer'' fields at the interface. As we shall see in section
\ref{theme3}, this does not appear to be necessarily possible for
front problems like streamers, which correspond to front propagation
into {\em unstable} states. The physical reason that the non-ionized
region into which the streamer fronts
propagate is unstable, is that as soon as there are free
electrons, there is further ionization due to the source term on the
right hand side of the streamer equations (\ref{stream1}),
(\ref{stream2}). This leads to an avalanch type of phenomenon, with
exponential growth of the electron density, characteristic of a
linearly unstable state.

Let us nonetheless not let ourselves get discouraged, but follow our
nose and assert that if the electron diffusion is small, one would
expect that the normal velocity of a streamer front like that of
Fig. \ref{dendretc}({\em c}) is approximately  equal to the drift velocity of the
electrons  on the outer side of the charge sheet, 
\begin{equation}\label{streamvn}
v_n \approx |\mu_e {\bf E}^+| = | \mu_e \nabla \Phi^+| ~.
\end{equation}
Here $\Phi $ is the electrical potential, ${\bf E}=-\nabla \Phi$, and
the superscript + indicates the value of the field at the charge
sheet, extrapolated from the non-ionized region. A linear relation,
like (\ref{streamvn}) is indeed found to a good approximation for
negatively charged streamers \cite{streamers}. Now, in the non-ionized
region outside the streamer, the charge density is essentially zero
[$n_+\approx 0, n_e\approx 0$ in (\ref{stream4})], and hence here
\begin{equation} \label{streamphi}
\nabla^2 \Phi \approx 0  \hspace*{1.2cm} \mbox{in the non-ionized region}~.
\end{equation}
Thus we see that if we do think of a streamer sheet as a moving charge
sheet and assume that the potential inside the streamer is roughly
constant due to the high mobility of the electrons, it falls within
the same class of gradient driven problems as 
dendrites and viscous fingers:  the normal velocity of the charge
sheet is proportional
to the gradient of a field $\Phi$, which itself obeys a Laplace or
diffusion equation.

\subsection{The Mullins-Sekerka instability \cite{langer,caroli,kassner}} 
We now discuss the Mullins-Sekerka instability of a planar interface.
We will first follow the standard analysis for the simplest case of a
planar solidification interface \cite{mullins,langer,caroli,kassner},
and then indicate why in the long wavelength limit the same
instability happens for all gradient-driven fronts whose outer field
obeys a Laplace or diffusion equation.  The analysis of the planar
interface appears at first sight to be somewhat of an academic
problem, as we will find such an interface to be unstable. However,
the analysis does identify the basic physics that is responsible for
the formation of natural growth shapes and it helps us to identify the
proper length scale for the growth shapes that result (the form of the
dispersion relation also plays a role in analytical approaches to the
dendrite   and viscous finger problem  \cite{reviews}).
\begin{figure}
\hspace*{3cm}
\epsfysize=3.5cm
\epsffile{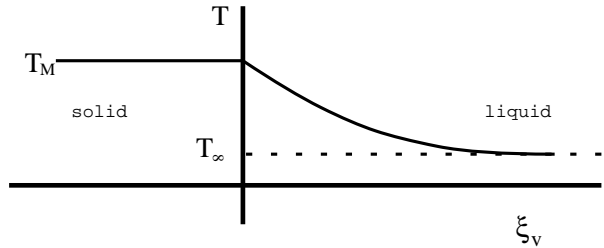}
\caption{\footnotesize\label{crystalgrowth}Qualitative sketch of the temperature
  profile at a planar interface in solidification.}
\end{figure}

We want to consider the stability of a planar interface which grows
with velocity $v$. To do so, we write the diffusion equation in a
frame $\xi_v=z-vt$ moving with velocity $v$ in the $z$-direction,
\begin{equation} \label{dendrdiff2}
  {{\partial T}\over{\partial t}} - v {{\partial T}\over{\partial
      \xi_v}}= D \nabla^2 T ~.
\end{equation}
Note that now $\nabla^2$ denotes the Laplacian in the moving
$x,y,\xi_v$ frame.
Furthermore, we consider for simplicity the limit $\beta \rightarrow
0$ so that the other two basic equations are Eqs. (\ref{dendrbc1}) and
(\ref{localeq}),
\begin{eqnarray}\label{dendrbc3}
  \hspace*{2.5cm} {{L}\over{c}} v_n & = & -D [ (\nabla T^\ell)_{n,int}
  - (\nabla T^s)_{n,int} ] ~,
\\ T^{\rm int} & = &  T_M [1 -(\sigma/L) \kappa]~.
\label{dendrbc4}
\end{eqnarray} 
Let us first look for a steady state solution, i.e., the solution for
a plane growing with a constant velocity $v$ in the $z$-direction into
an undercooled liquid. Since according to boundary condition
(\ref{dendrbc4}) $T^{\rm int}=T_M$ for a plane, the solution in the
solid is $T^s_0=T_M$, while solving the diffusion equation
(\ref{dendrdiff2}) for a solutions $T^\ell_0$ that are stationary in
the $\xi_v$ frame yields
\begin{equation}\label{plane}
  T^\ell_0 (\xi_v) = (T_M- T_\infty) e^{-\xi_v/\ell_D} +
  T_\infty ~,\hspace*{1cm} T^s_0=T_M~.
\end{equation}
Here we have taken the position of the plane at $\xi_v=0$,
$T_\infty$ is the temperature far in front of the plane, and $\ell_D
=D/v$ is the thermal diffusion length. The temperature profile given
by Eq. (\ref{plane}) is sketched in Fig. \ref{crystalgrowth}.
Substitution of this result into the boundary condition for heat
conservation (\ref{dendrbc3}) yields
\begin{equation}\label{undercooling}
  {{L}\over{c}} = (T_M-T_\infty)~.
\end{equation} 
This equation shows that the temperature $T_\infty $ has to be
precisely an amount $L/c$ below the melting temperature (this
criterion is often refered to as unit undercooling), and it is an
immediate consequence of heat conservation. For, in order for the
plane to be able to move with a constant speed, the amount of heat in
the diffusion boundary layer must be constant in time, in the
co-moving frame, and hence the net effect of the moving interface is
that it replaces a liquid volume element at a temperature $T_\infty$
by a solid volume element at a temperate $T_M$, while the heat per
unit volume which is generated is $L$. Equating this to the heat
$c(T_M-T_\infty)$ necessary to give the required increase in
temperature gives (\ref{undercooling}).

In passing, we note that if the undercooling far away is less
[$(T_M-T_\infty) <L/c$], a planar solidification front will gradually
slow down ($v\sim 1/\sqrt{t}$) due to the slow increase of the
thickness of the boundary layer. This gradual decrease of the speed is
slow enough that we can extend most of the analysis below by making a
quasistationary approximation for the velocity, but for simplicity we
will assume that condition (\ref{undercooling}) is satisfied.

We now turn to a linear stability analysis of this planar
interface. To do so, we assume that the interface is slightly
perturbed, i.e., that the position of the interface deviates slightly
from the planar position $\xi_v=0$. The strategy then is to write the
interface position as $\xi_v=\zeta(x,y,t)$ with $\zeta(x,y,t)$ small,
and to solve the diffusion equation and boundary conditions to first
order in $\zeta(x,y,t)$.
Since the unperturbed planar solution is translation invariant in the
$xy$ plane, the eigenmodes of the linearized equations are simple
Fouriermodes, and it suffices to analyze each Fourier mode separately.
Moreover, for simplicity we can take this mode to vary in the $x$
direction only. We thus write the perturbed interface and the
temperature field as single
Fourier modes of the form 
\begin{equation}\label{fourier}
  \zeta(x,y,t) = \zeta_k e^{\Omega t +ikx} ~,\hspace*{1cm} \delta T^{\ell,s} =
  \delta T^{\ell,s}(\xi_v) e^{\Omega t +ikx}~.
\end{equation}
Our goal is to determine the dispersion relation, i.e., $\Omega$ as a
function of $k$. If $\Omega$ is positive, the corresponding mode $k$
grows, and the planar solution is unstable to that perticular mode.
Consider first the temperature diffusion equation. Since it is already
linear, the functions $\delta T^{\ell,s}(\xi_v)$ satisfy the
simple differential equations
\begin{equation}\label{deltaT}
  {{d^2 \delta T^{\ell,s}_k(\xi_v)}\over{d\xi_v^2}} +
  {{1}\over{\ell_D}} {{d \delta
      T^{\ell,s}_k(\xi_v)} \over{d\xi_v }} = (\Omega/D
  -k^2)\delta T^{\ell,s}_k(\xi_v) ~.
\end{equation}
The solutions of these equations are simple exponentials; when we
impose that the perturbed temperature fields $\delta T^{\ell,s}$ have
to decay to zero far away from the interface, we get
\begin{eqnarray}\label{qs1} 
  \delta T^{\ell}_k(\xi_v) & =& \delta T^{\ell}_k
  e^{-q\xi_v}~,\hspace*{0.7cm} q={{1}\over{2 \ell_D}} \left( 1 +
  \sqrt{1-4\ell_D^2 \Omega/D + 4k^2\ell_D^2} \right)~,\\ \delta
  T^{s}_k(\xi_v)& =& \delta T^{s}_k e^{q'\xi_v}~,\hspace*{0.7cm}
  q'={{1}\over{2 \ell_D}} \left(-1 + \sqrt{1-4\ell_D^2 \Omega/D +
    4k^2\ell_D^2} \right)~.\label{qs2}
\end{eqnarray}
Furthermore, continuity of the temperature at the interface implies
$T^\ell_0 (\xi_v=\zeta)+ \delta T^\ell (\xi_v=\zeta) =T_M +
\delta T^s(\xi_v=\zeta)$. To linear order, we can take $\delta
T^\ell$ and $ \delta T^s$ at $\xi_v=0$, since they are already linear in
the perturbations. Expanding $T_0^\ell(\xi_v=\zeta)$ to linear
order gives $T_0^\ell(\xi_v=\zeta)
=T_M-(T_M-T_\infty)\zeta/\ell_D+..$ and so we simply get
\begin{equation}\label{inttemp}
  \delta T^\ell_k - (T_M-T_\infty)\ell^{-1}_D \zeta_k = \delta T^s~.
\end{equation}
Turning now to the boundary conditions (\ref{dendrbc3}) and
(\ref{dendrbc4}), we note that the curvature $\kappa$ of the surface
$\xi_v=\zeta$ becomes $\kappa = - {\partial^2 \zeta/\partial x^2
  }/[{(1+(\partial \zeta/\partial x)^2)^{3/2}}] = -\partial^2 \zeta
/\partial x^2 + O(\zeta^2)$. The local equilibrium interface boundary
condition (\ref{dendrbc4}) therefore becomes with this result and
(\ref{inttemp})
\begin{eqnarray}\label{Ts}
  \hspace*{2.5cm} \delta T^s_k & =& -(\sigma/L) T_M k^2 \zeta_k ~, \\ 
  \delta T^\ell_k & = & (T_M-T_\infty)\ell^{-1}_D \zeta_k -(\sigma/L)
  T_M k^2 \zeta_k~.
\end{eqnarray}
Finally, we need to linearize the conservation boundary condition
(\ref{dendrbc3}). The relation between the $z$-component of the
interface velocity and the normal velocity $v_n$ is
$v_z=v+\dot{\zeta}= v_n \cos \theta$, where $\theta$ is the angle
between the interface and the $z$ or $\xi$ direction. Since $\cos
\theta= 1/\sqrt{1+(\partial \zeta /\partial x)^2}$, this gives to
linear order $v_n=v + \dot{\zeta}$. Furthermore, the perturbed gradient
at the liquid side of the interface has two contributions, one from
$T_0^\ell$ evaluated at the perturbed position of the interface, and one
from $\delta T^\ell$. One gets, using also (\ref{undercooling}),
\begin{equation}\label{disp1}
  \Omega = {{v}\over{\ell_D}} \left[ -1 + q \ell_D + D (q+q')\left(
  -d_0 k^2 \right) \right]~,
\end{equation}
where
\begin{equation}
\label{d0}
d_0 = \frac{\sigma T_M c}{L^2}~,
\end{equation}
is the capillary parameter, which has a dimension of length. Just like
the ratio $\sigma/L$, $d_0 $ is typically a small microscopic length, of the order of tens of {\AA}ngstrom, say.

Eq. (\ref{disp1}) is the dispersion relation for the growth rate
$\Omega$ we were after. In this general form, it is not so easy to
analyze\footnote{It is easy to verify from 
  expressions (\ref{qs1}),(\ref{qs2}) that $\Omega=0$ for $k=0$. This
  is a consequence of the fact that the system is translation
  invariant, so that a perturbation that corresponds to a simple shift
  of the planar interface neither grows nor decays.} for general $k$,
since $q$ and 
$q'$ depend $k$ and $\Omega$ through Eqs. (\ref{qs1}) and (\ref{qs2}).

\begin{figure}
\epsfysize=4.2cm
\epsffile{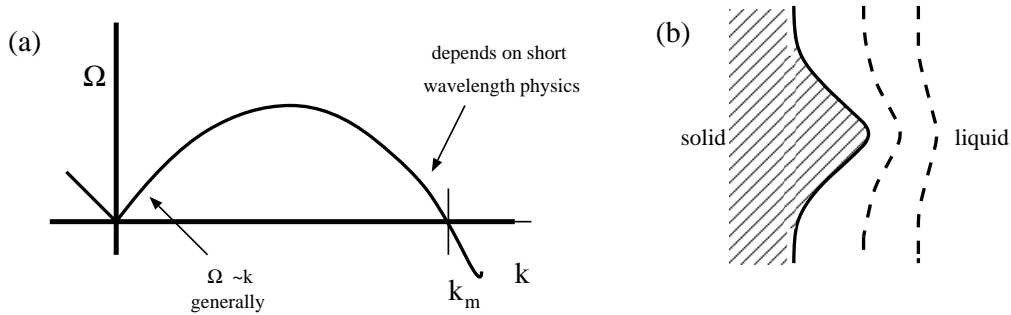}
\caption{\footnotesize\label{ms}{\em (a)} Sketch of the dispersion relation
  (\ref{disp2}) for the stability of the planar solidification
  interface, in the quasi-stationary approximation. The linear behavior
  of $\Omega$ with $|k|$ is generic for gradient driven growth
  problems, while the stabilization for larger $k$ values depends on
  the problem under consideration. {\em (b)}  Sketch of the
  compression of the isotherms in front of a bulge of the
  interface. If such a bulge appears on a long enough length scale
  that the capillary suppression of the local melting temperature is
  not too large, then the enhanced heat diffusion near the bulge
  associated with the compression of the isotherms makes the interface
  unstable. This is the origin of the Mullins-Sekerka instability,
  which is generic to gradient-driven growth problems.}
\end{figure}

The expression for $\Omega$ becomes much more transparent if the
diffusion coefficient $D$ is large enough and the perturbations of
short enough wavelength that both\footnote{One can not choose $\Omega$
  independently to satisfy these conditions; nevertheless, one can
  show that (\ref{disp2}) is a good approximation to (\ref{disp1}) if
  the diffusion coefficient is large enough that the conditions in the
  text are satisfied \cite{caroli}. Physically, we can think of this
  limit as the 
  one where the diffusion is so large that the temperature diffusion
  equation can be approximated by the Laplace equation.}
$\Omega \ll D k^2$ and $k\ell_D \ll 1$. This is actually the relevant
limit for small wavelengths, as then $\ell_D$ is very large and
timescales are slow. In this case, Eq. (\ref{qs1}) and (\ref{qs2})
show that $q'\approx q \approx |k|$, and then the dispersion relation
(\ref{disp1}) reduces to
\begin{equation}\label{disp2}
  \Omega \approx v |k| \left[ 1 -2 d_0 \ell_D k^2 \right]~.
\end{equation}
This is the form in which the dispersion relation is best known. As
Fig. \ref{ms} illustrates, $\Omega$ grows linearly for small $k$ (long
wavelength), and all modes with wave number $k < k_n=
1/\sqrt{2d_0\ell_D}$ have positive growth rates and hence are unstable
($k_n$ is the neutral wavenumber for which $\Omega=0$). Hence a mode
with this wavenumber neither grows nor decays). The maximum growth
rate is for $k_{max} =\sqrt{3}k_n$, i.e., for a wavelength
$\lambda_{max}=2\pi/k_n=2 \pi \sqrt{6d_0 \ell_D}$. We thus see that
the planar interface is unstable to modes within a whole range of
wavenumbers. Hence, even if we could prepare initially a (nearly) flat
interface, we would soon see that small protrusions, especially those
with a spatial scale of order $\lambda_{max}$, would start to grow
out. Quite soon, the interface evolution is then not described anymore
by the linearized equations, and one has to resort to some nonlinear
analysis to understand the morphology of the patterns that
subsequently arise. Typically, $\lambda_{max}$ still is an important
length scale even for these growth shapes, but it definitely is not
the only parameter that determines the scale and morphology of the
patterns  \cite{langer,caroli,kassner,reviews}. An example of this was
 discussed at the school by Brener  \cite{brener}.

Qualitatively, the origin of the Mullins-Sekerka instability is easy
to understand with the help of Fig. \ref{ms}{\em (b)}. If the
interface has some protrusion into the liquid, then the isotherms are
compressed in the neighborhood of this protrusion, provided the length
scale of the protrusion is large enough that the suppression of the
local melting term due to the capillary correction is small. This
means that the heat diffusion away from the interface is enhanced,
i.e., that the latent heat produced in this region of the interface
diffuses away more easily. Hence the interface can grow faster there,
and the protrusion grows larger in time.

It is important to realize that the instability that we identified
above only occurs upon growth, and not upon melting if the the heat
necessary to melt the crystal is supplied through the melt. This is
why ice cubes  keep a smooth rounded
shape during melting. You are encouraged to repeat the qualitative
arguments of Fig. 
\ref{ms}{\em (b)} to convince yourself of this\footnote{But nature
  always comes up with surprising exceptions: if a spin-polarized
  $^3He$ crystal 
  melts, a magnetic boundary layer builds up in the crystal, i.e., one
  now has a diffusion layer building up in front of a {\em melting}
  interface, while the temperature field is approximately homogeneous
  since the latent heat is small and the temperature diffusion fast. The Mullins-Sekerka instability upon melting that this
  results into due to the fact that both the interface velocity and
  the position of the diffusion boundary layer are reversed, was
  predicted in 1986  \cite{puech}. It  has just this
  summer been observed in the low temperature group in Leiden by
  Marchenkov {\em et al.}}. See \cite{langer} for help and further discussion of this point. 

Clearly, the physical mechanism underlying the Mullins-Sekerka
instability is not limited to crystal growth: it arises wherenever the
growth of or dynamics of a free interface is proportional to the
gradient of a field which itself obeys a Laplace equation or diffusion
equation --- in fact, the approximation $q\approx q' \approx k$ that
allowed us to reduce the dispersion relation to (\ref{disp2}) amounts
to replacing the diffusion equation by the Laplace equation in the
quasi-stationary limit! Now that we've done the analysis once in
detail, it is easy to see that the linear dispersion $\Omega \sim |k|$
which we found for solidification for small $k$ [See Fig. \ref{ms}{\em
  (a)}], is a general feature of diffusion limited or gradient driven
interface dynamics. To be specific, consider an interface whose normal
velocity $v_n$ is propertional to the gradient of some field $\Phi$,
which obeys the Laplace equation,
\begin{equation}
\label{laplace}
v_n = \nabla \Phi|_{int} ~,\hspace*{1.5cm} \nabla^2 \Phi =0
\hspace*{0.8cm} \mbox{in the bulk}~.
\end{equation}
For a planar solution with velocity $v$, we have then the solution
$\Phi_0(z)=\Phi'_0 z =v z$. Again, we consider perturbations
$\zeta=\zeta e^{\Omega t + ikx}$ of the interface. In order that
$\delta \Phi$ then obeys the Laplace equation, it must be of the form
\begin{equation}
  \delta \Phi =\delta \Phi_k e^{ikx-|k|z} ~, \hspace*{1.5cm} z>0~,
\end{equation}
Since to linear order the interface velocity $v_n$ in the presence of
the perturbation is $v+ \dot{\zeta}$, we now have
\begin{equation}\label{disp3}
  \Omega \zeta_k = -|k| \delta \Phi_k~.
\end{equation}
Finally the boundary conditions on the planar interface are such that
they can be written in terms of derivatives of the fields or the
interface shape, and if the basic equations are translation invariant
(i.e., there is no external field that tends to pin the position of
the interface), then we must have in linear order
\begin{equation}
  \Phi_0(\xi_v=\zeta_k) + \delta \Phi_k = ~~\mbox{terms of higher order
    in}~ k~,
\end{equation}
hence, since $\Phi_0'=v$,
\begin{equation}
  \delta \Phi_k =-v \zeta_k ~~\mbox{ as } k \rightarrow 0~.
\end{equation}
Using this in (\ref{disp3}), we find
\begin{equation}\label{disp4}
  \Omega \approx v |k| ~~\mbox{ as } k \rightarrow 0~.
\end{equation}
This clearly shows the generality of the presence of unstable long
wavelength modes with linear dispersion in gradient driven interface
dynamics. Not only solidification, but also viscous fingering,
streamer formation and flames \cite{comb1,comb2} are subject to this
same type of instability, as 
our discussion earlier in this section demonstrates. The differences
between the various problems mainly occur in the stabilizing behavior
at short distance scales. These depend on the details of the physics,
and are usually different for different problems. They have to be
included, however, since otherwise the interface would be completely
unstable in the short wavelength limit $k\rightarrow \infty$.

\subsection{The connection between viscous fingering and DLA}
An interesting illustration of the above observation is given by
Diffusion Limited Aggregation (DLA), in which clusters grow due to
accretion of brownian particles. Hence the driving force for growth is
essentially the same physics as above, a long-range diffusion field
governed by the Laplace equation, but in this case there are no
stabilizing smoothing terms at shorter wavelength. Only the particle
size or the lattice serves as a short distance cutoff, and in this
case the growth is fractal \cite{dla}.

The connection between the viscous fingering problem and DLA is
actually quite deep. In the viscous fingering case, the growth is
deterministic, and controlled by solving the Laplace equation in the
bulk. In DLA, the probability distribution of the random walkers is
also governed by the Laplace equation and the flux at the boundary of
the growing cluster is proportional to the gradient of the probability
distribution of walkers --- as we saw, this is the basic ingredient
 of the  Mullins-Sekerka instability. More importantly,
however, the DLA growth process is intrinsically noisy as one particle
is added at a time, and as there is no relaxation at the boundary of
the cluster. As pointed out by Kadanoff {\em et al.}
 \cite{kadanoff,bensimon}, the noise can be suppressed by having a
cluster grow only at a site once that site has been visited a number
of times by a random walker, and by allowing particles at the
perimeter to detach and re-attach to the cluster with a probability
that depends on the number of neighbors at each site. With increasing
noise reduction, DLA in a channel crosses over to viscous fingering.

Another surprising connection is that the mean occupation profile of
the average
of many realizations of DLA clusters in a channel approaches the shape
of a viscous finger. See  \cite{arneodo,dlavisc} for details.

\section{Smooth fronts as effective interfaces}\label{theme2}

\subsection{Fronts between a stable and a metastable state in one
  dimension --- existence, stability and relaxation}

We now turn our attention to a different but related issue, namely the
question when we can map a model with a smooth front, domain wall or
transition zone, onto a sharp interface model, with boundary
conditions which are local in space and time.  The answer to this
question, namely that this typically can be done for problems in which
the interface separates two (meta)stable states or phases may not be
that surprising. Nevertheless, thinking about these issues helped us
clarify some of the points which we feel have not been paid 
due attention to in the literature, and which come to the foreground
in our work with Ebert and Caroli on streamers  \cite{streamers}. There
you really run into trouble if you blindly apply the formalism as it
is usually presented in the literature. This will be discussed in
detail in our future publications with Ebert \cite{ebert2}, and I will
keep you in suspense till  section \ref{theme3} for a brief
sketch of our present results and implications. Further motivation for
the analysis of this section was given in the introduction.

I am convinced most --- if not all --- elements of the discussion
below must appear at many places in the literature. For example, the
first part of the analysis appears in one form or another in
 \cite{gunton,bray,langer2,allen}, but since it will arise in almost
any Ginzburg-Landau type of analysis --- the working horse of
condensed matter physics --- I presume most ingredients can be found
at many more places (similar questions arise in the analysis of
instantons in field theory). Nonetheless, we have not come across any
discussion from the perspective that we will emphasize in
\cite{ebert2}, the 
relation between relaxation, interface limits, and solvability. The
present section is intended to  provide
a summary of the background material that can be found at scattered
places in the literature and to serve as an introduction our papers
\cite{ebert2}.

To be concrete, we will present our discussion in terms of a dynamical
equation in one dimension of the form
\begin{equation}\label{op1}
  {{\partial \phi}\over{\partial t}}= {{\partial^2 \phi}\over{\partial
      x^2}} + g(\phi)~.
\end{equation}
Here $\phi$ is a real order parameter. This equation is about the
simplest model equation for the analysis of relaxation dynamics, but
it captures the essentials of the issues that also arise in more
complicated variants and extensions.

Later, in our discussion of the coupling to other fields, it is useful
to introduce appropriate parameters to tune the time and spatial
scales of the variation of $\phi$, but for the present discussion of
Eq. (\ref{op1}) we will not need these. We have therefore used the
freedom to choose appropriate time and spatial scales to set the
prefactors of the derivative terms to unity.

It will turn out to be useful to express $g(\phi)$ in terms of the
derivative of two other functions, which both play the role of a
potential in different circumstances:
\begin{equation}\label{g}
  g(\phi) \equiv - {{d f(\phi)}\over{d\phi }} \equiv {{d
      V(\phi)}\over{d\phi }} ~,
\end{equation}
so that equivalent forms of (\ref{op1}) are
\begin{equation}\label{op2}
  {{\partial \phi}\over{\partial t}}= {{\partial^2 \phi}\over{\partial
      x^2}} - {{d f(\phi)}\over{d\phi }} \hspace*{1.5cm}
  \Longleftrightarrow \hspace*{1.5cm} {{\partial \phi}\over{\partial
      t}}= {{\partial^2 \phi}\over{\partial x^2}} + {{d
      V(\phi)}\over{d\phi }}~.
\end{equation}
As we shall see later on, $f$ has the interpretation of a free energy
density in a Ginzburg-Landau picture, while $V$ will play the role of
a particle potential  in a standard argument in which there is a
one-to-one correspondence between front solutions and trajectories of
a particle moving in the potential $V$. 

\begin{figure}
\epsfxsize=13.6cm
\epsffile{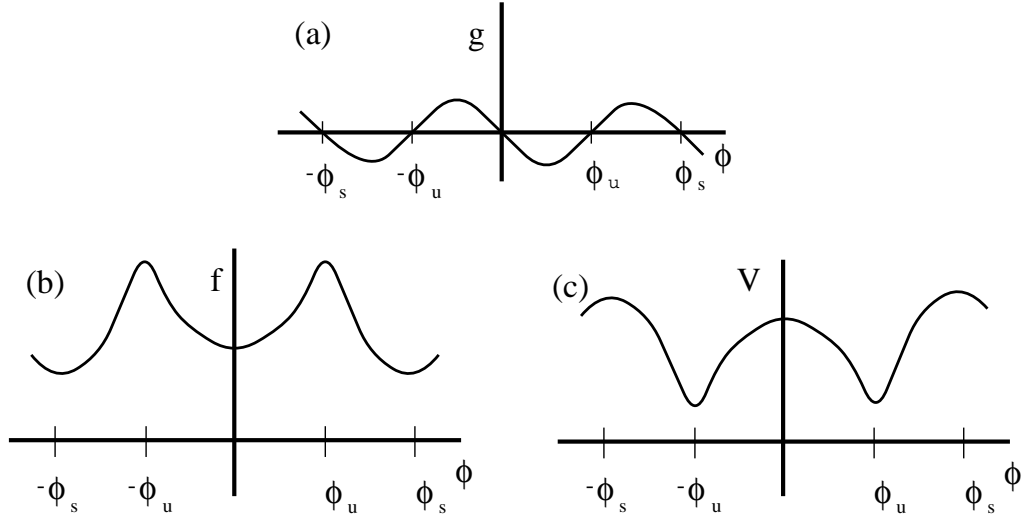}  
\caption{\footnotesize\label{gfv}The function $g(\phi)$ and the associated
  potentials $f(\phi)$ and $V(\phi)$ used in our discussion of front
  solutions of Eq. (\ref{op1})}
\end{figure}

We are interested in cases in which $g(\phi)$ has two zeroes
$g(\phi_s)$ with $g'(\phi_s)<0$; these correspond to (meta)stable
homogeneous solutions $\phi=\phi_s$ of Eq. (\ref{op1}). Indeed, if we
linearize this equation about $\phi_s$, and substitute $\Delta \phi
\sim e^{\Omega t + ikx}$ (very much like we've done before), then we
find $\Omega = g'(\phi_s)- k^2 <0$, which confirms that the state is
linearly stable. Without loss of generality, we can always take
$g(0)=0$ ($V'(0)=0$), so that $\phi=0$ is one of the stable
states. We will label the other linearly stable state simply $\phi_s$.
Although this is not necessary, we will for simplicity also take $g$
antisymmetric [$g(-\phi)=g(\phi)$], so that the potentials $f$ and $V$
are symmetric. A typical example of a function $g(\phi)$ and its
corresponding potentials is sketched in Fig. \ref{gfv}.  Note that
there is also a third root of $g(\phi)$ in between 0 and $\phi_s$, and
that here $g'(\phi_u)>0$. A homogeneous state $\phi=\phi_u$ is
therefore unstable ($\Omega=g'-k^2 >0$ for small $k$).

\begin{figure}
\epsfxsize=13.6cm
\epsffile{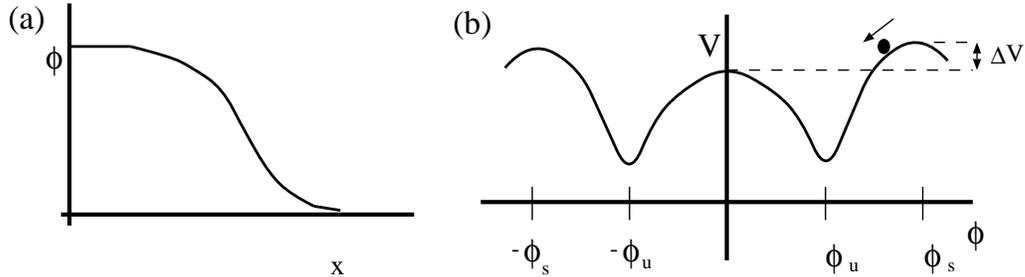}
\caption{\footnotesize\label{phiv}{\em (a)} Example of the type of moving front
  solution we are looking for. {\em (b)} The potential $V$. A moving
  front solution like the one sketched in {\em (a)} corresponds in the
  particle-on-the-hill analogy to the solution of the dynamical
  problem in which the particle starts at the top at $\phi_s$, moves
  down the hill and up the one in the center, and comes to rest at the
  center top as the quasi-time $\xi_v \rightarrow \infty$.}
\end{figure}

Let us now focuss right away on front or domain wall type solutions of
the type sketched in Fig. \ref{phiv}{\em (a)}: they connect a domain
where $\phi\approx \phi_s$ on the left to a domain where $\phi\approx
0$ on the right. Obvious questions are: what  does the solution look
like? In which direction will the front move? And how does it relax to
its moving state? The answers to these questions can be obtained in a
very appealing and intuitive way for this simple model equation by
reformulating the questions into a form that almost every physicist
is familiar with. However, the two main points --- the existence of a
unique solution and exponential relaxation --- have more general
validity.

We can look for the existence of moving front solutions by making the
Ansatz $\phi_v(x-vt) =\phi_v(\xi_v)$, with $\xi_v=x-vt$. Such solutions
are uniformly translating in the $x$ frame, and hence stationary in
the co-moving  frame $\xi_v$.  Substitution of this Ansatz into Eq.
(\ref{op2}) gives, after a rearrangement
\begin{equation}\label{particle}
  {{d^2 \phi_v}\over{d \xi_v^2}}= - v {{d \phi_v}\over{d \xi_v}} - {{d
      V(\phi_v)}\over{d\phi_v }}~.
\end{equation}
This equation is familiar to you: it is formally equivalent to the
equation for a ``particle'' with mass 1 moving in a potential $V$, in
the presence of ``friction''. In this analogy, which is summarized
below, $\xi_v$ plays the role of time, and $v$ the role of a friction
coefficient:
\begin{equation}
\begin{array}{cccc}\label{analogy}
  \xi_v & \phi_v & v & V \\ \Updownarrow &\Updownarrow &\Updownarrow
  &\Updownarrow \\ \hspace*{0.4cm} \mbox{time} \hspace*{0.4cm}
  &\hspace*{0.4cm} \mbox{displacement}\hspace*{0.4cm} &\hspace*{0.4cm}
  \mbox{friction coefficient}\hspace*{0.4cm}
  &\hspace*{0.4cm}\mbox{potential} \hspace*{0.4cm}
\end{array}
\end{equation}
Clearly, the question whether there is a traveling wave solution of
the type sketched in Fig. \ref{phiv}{\em (a)} translates into the
question, in the particle-on-the-hill analogy: is there a solution in
which the particle starts at the top of the potential $V$ at $\phi_s$
at ``time'' $\xi_v=- \infty$, rolls down the hill, and comes to rest at
the top of the hill at $\phi=0$? In the language of the analogy the
answer is immediately obvious: if the value $V(\phi_s)$ of the
potential at $\phi_s$ is larger than at $\phi=0$, i.e., if
\begin{equation}\label{deltaV}
  \Delta V \equiv V(\phi_s) - V(0)
\end{equation}
is positive, then there must be a solution with a nonzero positive
value of the velocity (the ``friction coefficient''). Such a solution
corresponds to a front which moves to the right so that the
$\phi\approx \phi_s$ domain expands. In the opposite case, when the
potential at $\phi_s$ is lower than at 0 so that $\Delta V<0$, then
such a solution only exists for ``negative friction'' so that enough
energy is pumped into the system that the particle can climb the center hill.
Negative friction corresponds to a left-moving front with $v<0$, so
that the $\phi\approx 0$ domain expands.

\begin{figure}
\epsfxsize=10cm
\hspace*{1.8cm}
\epsffile{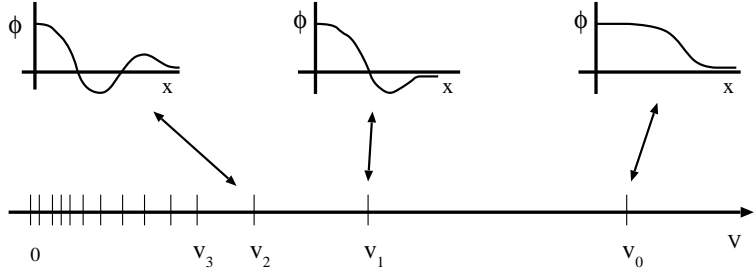}
\caption{\footnotesize\label{vis}Graphical representation of the fact that  a
  discrete set of moving front solutions is found at velocity values
  $v_0,v_1,v_2,\ldots$. The number of nodes of the corresponding
  profiles $\phi_{v_i}$ is $i$.}
\end{figure}

Let us make this a bit more precise by first asking what happens when
the 
``friction'' $v$ is very large. Then, there is no solution where the
particle moves from the top at $\phi_s$ to the one at $\phi=0$: for
large friction the particle creeps down the hill and comes to rest in
the bottom of the potential. When the friction is reduced, the
particle looses less energy, and is able to climb further up on the
left side of the well. Hence if we keep on reducing the ``friction'' $v$,
at some value $v=v_0$ the particle has just enough energy left to
climb up all the way to the top at $\phi=0$, and get to rest there. In
other words, at $v=v_0$, there is a unique solution of the type
sketched in Fig. \ref{phiv}({\em a}). If $v$ is reduced slightly
below $v_0$, the particle overshoots a little bit, and it finally ends
up in the left well. So for $v$ just below $v_0$, there are no
solutions with $\phi_v \rightarrow 0$ for $\xi_v\rightarrow 0$. However,
if we keep on reducing $v$, there comes a point $v=v_1$ where the
particle first overshoots the middle top, then moves   back and
forth once in the left well, and finally makes it to the center top
--- the 
profile $\phi_{v_1}(\xi_v)$ then has one node where
$\phi_{v_1}(\xi_v)=0$. Clearly, 
we can continue to reduce $v$ and find values $v_2,v_3,v_4,\ldots$
where the profile $\phi_{v_i}$ has $2,3,4,\ldots$ nodes. As is
illustrated in Fig. \ref{vis}, we thus have a discrete set of moving
front solutions. Which one is stable and dynamically relevant?
Intuitively, we may expect that the one with the largest velocity,
$v_0$, is both the stable and the dynamically relevant one, since the
multiple oscillations of the other profiles look rather unphysical.
This indeed turns out to be the case: if you start with an initial
condition close to the profile $\phi_{v_1}$ with velocity $v_1$, you
will find that the node either ``peels off'' from the front region and
then stays behind, or moves quickly ahead to disappear from the scene
on the front end. In both cases, a front with velocity $v_0$ emerges
after a while. The stability analysis of the front solutions which we
will present later confirms that only the fastest $v_0$ front solution
is linearly stable.

Before turning to the linear stability analysis, we make a brief
digression about the connection with a more thermodynamic point of
view that is especially popular in studies of
coarsening \cite{gunton,bray,langer2,allen}.

It is well known that Eqs. (\ref{op1}),(\ref{op2}) can also be written
as
\begin{equation}
\label{gl}
{{\partial \phi}\over{\partial t}}= - {{\delta F}\over{\delta \phi}}~,
\hspace*{1.6cm} F= \int dx \left[ \half \left( \frac{\partial
  \phi}{\partial x}\right)^2 + f(\phi) \right]~.
\end{equation}
In a Ginzburg-Landau like point of view, $F$ plays the role of a free
energy functional, whose derivative $\delta F/\delta \phi$ drives the
dynamics, and $f(\phi)$ is the coarse grained free energy
density. This formulation brings out clearly that the dynamics is 
relaxational and corresponds to that of a non-conserved order
parameter (the conserved case corresponds to $\partial \phi / \partial
t = + \nabla^2 \delta F / \delta \phi$, so that $\int dx~ \phi $
remains constant under the dynamics). Note also that in statistical
physics one often starts with postulating an expression for a coarse
grained free energy functional like $F$, and then obtains the dynamics
for $\phi$ from the first equation in (\ref{gl}). One should be aware
that in pattern formation, one usually has to start from the dynamical
equations, and that these usually do not follow from some simple free
energy functional  \cite{sanmiguel}.

An immediate consequence of (\ref{gl}) is that
\begin{equation}
  {{d F}\over{dt}} = \int dx~ {{\delta F}\over{\delta \phi }}
  {{\partial \phi }\over{\partial t}} = - \int dx \left( {{\delta
      F}\over{\delta \phi }}\right)^2 \leq 0~,
\end{equation}
so that under the dynamics of $\phi$, $F$ is a non-increasing function
of time --- it either decreases or stays constant (in technical terms:
$F$ is a Lyapunov functional). Since the homogeneous steady states
$\phi=0$ and $\phi=\phi_s$ correspond to minima of $f(\phi)$ and hence
of $F$, this immediately shows that a front moves in the direction so
that the domain whose  state has the lowest free energy density $f$ expands. Since
$f=-V$, this is equivalent to the conclusion reached above, that the
domain corresponding to the maximum value of the potential $V$ expands.

Consider now the case in which the states $\phi=0$ and $\phi_s$ have
the same free energy density: $\Delta f \equiv f(\phi_s) - f(0)=
-\Delta V=0$. They are then ``in equilibrium'', and a wall or
interface between these two states does not move.  Then the excess
free energy per unit area, associated with the presence of this wall,
which is nothing but the {\em surface tension} $\sigma$ is
\begin{equation}\label{sigma0}
  \sigma = \int dx \left[ \half \left( \frac{\partial \phi_0}{\partial
    x}\right)^2 + f(\phi_0) - f(0) \right]~.
\end{equation}
We can rewrite this by using the fact that energy conservation in the
particle picture implies that the sum of the kinetic and ``potential''
energy ($-f$) is constant, so that $\half (\partial \phi_0 / \partial x)^2
-f(\phi_0) = -f(0)$, since far away to the right $\partial \phi_0 /
\partial x \rightarrow 0$ and $\phi_0 \rightarrow 0$. Using this in (\ref{sigma0}), we get
\begin{equation}\label{sigma}
  \sigma = \int dx \left( \frac{\partial \phi_0}{\partial
    x}\right)^2~,
\end{equation}
which is an expression which is very often used in square-gradient
theories of interfaces. In our case, we can use it to obtain a
physically transparant expression for the velocity $v$ of the moving
front: If we multiply (\ref{particle}) by $d\phi_v/d\xi_v$ and integrate
over $\xi_v$, the term on the left side becomes $\int d\xi_v (d\phi_v
/d\xi_v) (d^2\phi_v/d\xi_v^2)$ $= \half \int d\xi_v (\d/d\xi_v) (d\phi_v/d\xi_v)^2=0$
since $d\phi_v/d\xi_v$ vanishes for $\xi_v \rightarrow \pm \infty$. As a
result, we are left with
\begin{equation}\label{vexpr}
  v = {{\int d\xi_v \left[ {{d \phi_v}\over{d\xi_v}} {{df}\over{d\phi}}
    \right] }\over{\int d\xi_v \left( {{d \phi_v}\over{d\xi_v}} \right)^2
    }} = {{ - \Delta f }\over{\int d\xi_v \left( {{d \phi_v}\over{d\xi_v}}
  \right)^2 }}~.\label{vlinear}
\end{equation}
This expression confirms again that the domain whose state has the
lowest free energy $f$ expands. But it shows more: for small
differences $\Delta f$, the velocity is small, so we can approximate
$\phi_v$ in the denominator by $\phi_0$, the profile of the interface
in equilibrium. But in this approximation, the denominator is nothing
but the surface tension of Eq. (\ref{sigma}), so
\begin{equation}
  v\approx {{- \Delta f}\over{\sigma}}~, \hspace*{1.5cm}v~\mbox{small}~.
\end{equation}
Thus, the response of the interface is linear in the driving force
$\Delta f$ and the surface tension $\sigma$ plays the role of an
inverse mobility coefficient. The above expressions are often used in
the work on coarsening, and can be extended to include perturbatively
the effect of curvature or slowly varying additional fields on the
interface velocity. We will come to this later.

We now return to the question of stability of the front solutions with
velocity $v_0,v_1,\ldots$, using an analysis that is inspired by a few
simple arguments in  \cite{benjacob}. Keep in mind that we will study
the stability of front solutions in one dimension themselves, {\em
  not} the stability of a planar interface or front to small changes
in its shape, like we did in section \ref{theme1}. To study the linear
stability of a front solution $\phi_v(\xi_v)$, we write
\begin{equation}\label{eta}
  \phi_{v_i}(\xi_v,t)=\phi_{v_i}(\xi_v) + \eta(\xi_v,t)~,
\end{equation}
and linearize the dynamical equation (\ref{op1}) in $\eta$ in the
moving frame $\xi_v$ to get
\begin{equation}\label{eta2}
  {{\partial \eta}\over{\partial t}}= v_i {{\partial \eta}\over{\partial
      \xi_v}} + {{\partial^2 \phi}\over{\partial x^2}} + g'(\phi_{v_i})
  \eta + O(\eta^2)~.
\end{equation}
Since the equation is linear, we can answer the question of stability
by studying the spectrum of temporal eigenvalues. To do so, we write
\begin{equation}\label{psi}
  \eta(\xi_v,t) = e^{-Et} e^{-v\xi_{v} /2} \psi_E(\xi_v)~.
\end{equation}
so that all modes with eigenvalues $E>0$ are stable.  Upon
substitution of this in (\ref{eta2}), we get
\begin{equation}  \begin{array}[t]{r}
  \left[ - {{\partial^2}\over{\partial \xi_v^2}} + \underbrace {\left(
      {{v^2}\over{4}} -g'(\phi_{v_i}) \right)}\right] \\ U(\xi_v)
    \hspace*{1.2cm} \end{array} \psi_E(\xi_v) = E \psi_E (\xi_v)~,
\end{equation}
which is nothing but the Schr\"odinger equation (with $\hbar^2/m =1$)
and which explains why we used $E$ for the temporal eigenvalue in the
Ansatz (\ref{psi}). In the analogy with quantum mechanics which we
will now exploit, $U(\xi_v)$ plays the role of a potential , and we are
interested in the energy eigenvalues $E$ of the quantum mechanical
particle in this potential. If we find a negative eigenvalue, the
profile $\phi_{v_i}$ is unstable, i.e., there is then at least one
eigenmode of the linear evolution operator whose amplitude will grow
in time under the dynamics.  In 
other words, if we take as an initial condition for the dynamics the
uniformly translating profile $\phi_v$ we considered plus a small
perturbation about this which has a decomposition along this unstable
eigenmode, the perturbation proportional to this eigenmode will grow
in time. 
\begin{figure}
\epsfxsize=12.5cm
\hspace*{0.5cm}
\epsffile{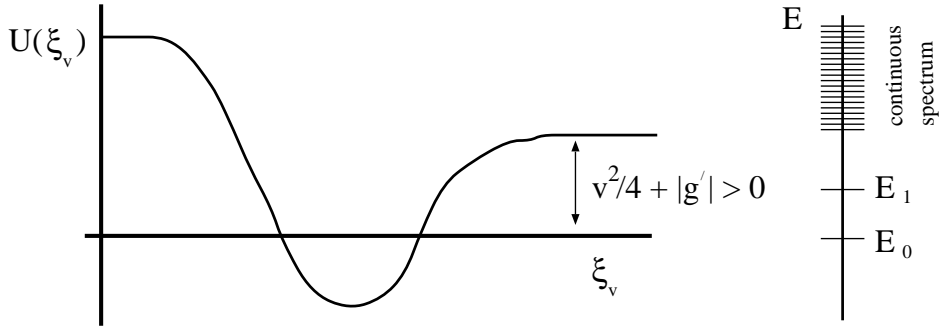}
\caption{\footnotesize\label{U}Sketch of the potential $U(\xi_v)$ which enters in the
  stability analysis of the front $\phi_{v_0}$ between a stable and a
  meta-stable state. The asymptotes $U(\infty)$ and $U(-\infty)$ are
  both positive. The resulting eigenvalues spectrum is sketched on the
  right. Note that there always is an eigenvalue $E=0$ due to the
  translation mode. }
\end{figure}

Consider first the form of the potential $U(\xi_v)$ for $v=v_0$. In this
case the front has a smooth monotonically decreasing profile of the
form sketched in Fig. \ref{phiv}({\em a}). Both for $\xi_v \rightarrow -
\infty$ and for $\xi_v \rightarrow \infty$, $g'(\phi_{v_0})$ is
negative, so $U(\xi_v)$ is positive for $\xi_v \pm \infty$. In between,
around $\phi_{v_0}=\phi_u$, $g'(\phi)$ is positive as Fig. \ref{gfv}{\em (a)}
shows, and so $U(\xi_v)$ is smaller than its asymptotic values in this
range. The resulting shape of $U(\xi_v)$ is sketched in Fig. \ref{U} for
a case in which $U(-\infty)>U(\infty)$. Armed with a physicist's
standard knowledge of quantum mechanics, we can now immediately draw
the following conclusions:\\ \hspace*{1cm} {\em (i)} The continuous
spectrum corresponds to solutions $\psi_E$ that approach plane wave
states as $\xi_v\rightarrow \infty$ in the case drawn in Fig. \ref{U},
and so they have an energy $E \geq U(\infty)=(v_0^2/4 +|g'(0)|) >0$.
In other words, the bottom of the continuous spectrum lies at a
positive 
energy, and all the corresponding eigenmodes relax exponentially
fast.\\ \hspace*{1cm} {\em (ii)} Next, consider the discrete spectrum.
Since the original equation is translation invariant, if
$\phi_{v_0}(\xi_v) $ is a solution, so is
$\phi_{v_0}(\xi_v+a)=\phi_{v_0}(\xi_v)+ a ~d\phi_{v_0}(\xi_v)/d\xi_v+\cdots$.
In other words, as the perturbation is nothing but a small
shift of the profile, the perturbation should neither grow nor decay.
This implies that $d\phi_{v_0}(\xi_v)/d\xi_v$ must be a ``zero mode'' of
the linear equation, i.e., be a solution of the Schr\"odinger equation
(\ref{psi}) with eigenvalue $E=0$:\footnote{You can easily convince
  yourself that this is true by substituting $\phi_{v_0}(\xi_v+a)$ in
  the original ordinary differential equation for the profile
  (\ref{particle}), expanding to linear order in $a$, and transforming
  to the function $\psi$. You then get (\ref{psi}) with $E=0$ and
  $\psi_0=d\phi_{v_0}(\xi_v)/d\xi_v$.}
\begin{equation}
  E=0: \hspace*{1cm} \psi_0(\xi_v) = {{d \phi_{v_0}(\xi_v)}\over{d\xi_v}} ~.
\end{equation}
\hspace*{1cm} {\em (iii)} Clearly, the ``translation mode''
$d\phi_{v_0}(\xi_v)/d\xi_v$ with eigenvalue zero is a ``bound state
solution'' [as it should, in view of {\em (i)}] since it decays
exponentially to zero for $\xi_v \rightarrow \pm \infty$. Moreover,
since $\phi_{v_0}$ decays monotonically, $d\phi_{v_0}(\xi_v)/d\xi_v<0$, so
the translation mode $d\phi_{v_0}(\xi_v)/d\xi_v$ does not have a zero,
i.e., is nodeless. Now it is a well-known result of quantum
mechanics \cite{messiah} that the bound state wave functions can be
ordered according to the number of nodes they have: the ground state
with energy $E_0$ has no nodes, the first excited bound state (if it
exists) has one node, and so on. If we combine this with our
observation that $\psi_0=d\phi_{v_0}(\xi_v)/d\xi_v$ is nodeless and has an
eigenvalue $E_0=0$, we are led immediately to the conclusion that if
there are other bound states, they {\em must} have eigenvalues $E>0$.

Taken together, these results show that apart from the trivial
translation mode all
eigenfunctions\footnote{\label{footnotewarning}There is actually a
  slightly subtle issue here that we have swept under the rug. In
  quantum mechanics, wave functions $\psi$ which diverge as $\xi_v
  \rightarrow \pm \infty$ are excluded, as these can not be
  normalized; due to the 
  transformation (\ref{psi}) from $\eta$ to $\psi$, there can be
  perfectly honorable eigenfunctions $\eta$ of fronts that do not
  translate into normalizeable wave functions $\psi$. In the present
  case, these eigenfunctions turn out to have large positive eigenvalues, and
  so they do not affect our conclusions concerning the relaxation, but
  for fronts propagating into unstable states one has to be much more
  careful. See section \ref{theme3} and  \cite{ebert2} for further
  details.} have positive 
eigenvalues $E$ and so are stable: they decay as $t \rightarrow
\infty$. Moreover, there is a gap: if the form of the function
$g(\phi)$ is such that there are bound state solutions, then the mode
that relaxes slowest is the first ``excited'' bound state solution
$\psi_1$ with eigenvalue $E_1>0$. Otherwise, the slowest relaxation
mode is determined by the bottom of the continuous spectrum. In either
case, {\em all nontrivial perturbations around the profile
  $\phi_{v_0}$ relax exponentially fast}.

It is now easy to extend the analysis to the other front profile
solutions $\phi_{v_1}, \phi_{v_2}$, etc.  Consider, e.g.,
$\phi_{v_1}$. The analysis of the continuous spectrum proceeds as
before so the continuous spectrum again has a gap. Again, the
translation mode $d\phi_{v_1}(\xi_v)/d\xi_v$ neither grows nor decays, so
has eigenvalue zero, but now the fact that $\phi_{v_1}$ goes through
zero once and then decays to zero implies that $d\phi_{v_1}(\xi_v)/d\xi_v$
has exactly one node. According to the connection between the number
of nodes of bound state solutions and the ordering of the energy
eigenvalues, there must then be precisely one eigenfunction with a
smaller eigenvalue than the translation mode which has $E=0$. In other
words, there is precisely one unstable mode. Likewise, all other
profiles $\phi_{v_i}$ with $i\geq 0$ are unstable to $i$ modes.

In summary, our analysis shows that in the dynamical equation
(\ref{op1}) for the order parameter $\phi$, there is a discrete set of
moving front solutions. Only the fastest one is stable, and its motion
is in accord with simple thermodynamic intuition. Moreover, the
relaxation towards this unique solution is exponentially fast, as
$e^{-\Delta E t}$, where $\Delta E$ is the gap to the lowest bound
state eigenvalue, if one exists, or else to the bottom of the
continuum band.

\subsection{Relaxation and the effective interface approximation}
As explained in the introduction, in many cases one wants to map a
problem with a smooth but thin front or interfacial zone onto one with
a mathematically sharp interface with appropriate boundary conditions.
We have termed this the {\em effective interface approximation}.
Reasons for using this mapping can be either to replace a sharp
interface problem by a computationally simpler one with a smooth front
(e.g., a so-called phase-field model for a solidification front
\cite{fife,karma,kupferman}) or to translate a problem with a thin
transition zone (e.g., streamers \cite{streamers}, chemical waves
\cite{meron,goldstein2}, combustion \cite{comb1,comb2}) onto a moving
boundary problem, so as to be able to exploit our understanding of
this class of problems. We will refer to this literature and to
\cite{fife2} for detailed discussion of the mathematical basis of such
approaches.  Here, we just want to emphasize how the exponential
relaxation of front profiles that we discussed above is a {\em
  conditio sine qua non} for being able to apply this mapping.

For concreteness, let us consider the following phase-field model
which is a simple example of the type of models which have been
introduced for studying solidification within this context
\begin{eqnarray}\label{pp1}
  \hspace*{2.6cm} {{\partial u}\over{\partial t}} & = & \nabla^2 u +
  {{\partial \phi}\over{\partial t}} ~, \\ \varepsilon {{\partial
      \phi}\over{\partial t}} & = & \varepsilon^2 \nabla^2 \phi +
  g(\phi,u) ~,\label{pp2} \\ g(\phi,u) &=& - {{\partial
      f}\over{\partial 
      \phi}} ~, \hspace*{1cm} f(\phi,u) = \phi^2 (\phi -1)^2 + \lambda
  u 
  \phi ~.\label{pp3}
\end{eqnarray}
In this formulation, $\phi$ is the order parameter field, and $u$
plays the role of a temperature. For fixed $u$, we recognize in
(\ref{pp2}) the order parameter equation that we have studied before:
the potential $f$ has a double well structure for $\lambda u$ small.
At $u=0$ the states $\phi=0 $ and $\phi=1$ have the same free energy
$f$, and the ``liquid'' state $\phi=0$ and ``solid'' state $\phi=1$ are
then in equilibrium. As we have seen, an interface beween these two
states then neither melts nor grows. For $\lambda>0$, a positive
temperature $u$ makes the liquid-like state at the minimum near
$\phi=0$ the lowest free energy state, and below the melting
temperature $u=0$ the solid-like minimum near $\phi=1$ has the lowest
free energy. The order parameter equation is coupled to the diffusion
equation (\ref{pp1}) for the temperature through the term $\partial
\phi /\partial t$. This term plays the role of a latent heat term when
solidification occurs: it is a source term in the interfacial zone,
where $\phi$ rapidly increases from about zero to one. Moreover, if
the interface is locally moving with speed $v_n$, then $\partial \phi
/ \partial t \approx -v \partial \phi /
\partial \xi_v$, so if we integrate through the thin interfacial zone we
see that this term contributes a factor $v_n$, in agreement with the
the fact that the latent heat released at the solid-melt interface is
proportional to $v_n$.

In writing Eqs. (\ref{pp1})-(\ref{pp3}), the space and time scales
have been written in units of the ``outer'' scale on which the
temperature field $u$ varies. In these units, the interface width in
the order parameter field $\phi$ should be small, and this is why the
parameter $\varepsilon \ll 1$ has been introduced in (\ref{pp2}): it
ensures that the interface width $W$ scales as $\varepsilon$ and that
the time scale $\tau$ for the order parameter relaxation is also of
order $\varepsilon$. It thus allows us to derive the effective
interface equations mathematically using the methods of matched asymptotic
expansions or singular perturbation theory  \cite{fife,bender,vandyke}
by taking the limit $\varepsilon \rightarrow 0$. Since both $W$ and $\tau$
scale as $\varepsilon$ the response of the interface velocity $v_n$
stays finite as $\varepsilon$ goes to zero.
\begin{figure}
\epsfxsize=9cm
\hspace*{2.5cm}
\epsffile{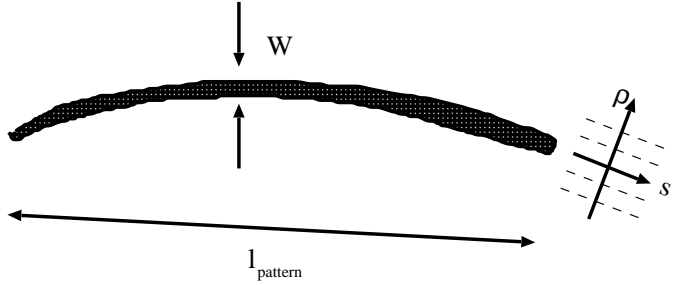}
\caption{\footnotesize\label{curvedfront}Qualitative sketch of a curved front of
  width $W$, and the local curvilinear coordinate system $(\rho,{\bf
    s})$ used in the derivation of an effective interface model.}
\end{figure}

Although the mathematical analysis by which effective interface
equations can be obtained is certainly more sophisticated and
systematical than what will transpire from the brief discussion in
this section, what seems to be the essential step in all the
approaches is the following. In the term $\lambda u$ in $g$ or $f$,
which is often treated for convenience as a small perturbation,  it is
recognized that in the 
interfacial zone (of width of order $\varepsilon$) $u$ does not change
much and hence can effectively be treated as a constant in lowest
order. Moreover, 
since the shape of the interface is curved on the ``outer'' scale, the
curvature $\kappa$ of the interfacial zone, when viewed on the inner
scale of the front width, is treated as a small parameter which
enters, as we shall see below, the equations in order
$\varepsilon$. This is because when $W\rightarrow 0$, the front
becomes locally almost planar. As is illustrated in
Fig. \ref{curvedfront}, 
one now introduces a curved local coordinate system $\rho({\bf r},t),
{\bf s({\bf r},t)}$ where the $\rho$ is oriented normal to the front
and points in the direction of the  $\phi \approx 0$ phase, which in a
Ginzburg-Landau description is normally associated with the disordered
phase (we thought of it as the ``liquid'' phase before). 
By choosing, e.g., the line $\rho=0$ to coincide with the contour line
$\phi = \half$, we ensure that this line follows the interface zone.
In the limit $\rho\rightarrow 0$ we then have
\begin{equation}
  \lim_{ \rho \rightarrow 0 } \left. {{\partial \rho }\over{\partial
      t}} \right|_{\bf r} = -v_n ({\bf s},t)~, \hspace*{1cm} \lim_{ \rho
    \rightarrow 0} \nabla^2 \rho = \kappa ({\bf s},t)~.
  \label{vandkappa}
\end{equation}
The derivation of an effective interface approximation now proceeds by
introducing the stretched (curvilinear) coordinate $\xi_v=\rho /
\varepsilon$ for the analysis of the inner structure of the front
profile, and assuming that the fields $\phi$ and $u$ can be expanded
in a power series of $\varepsilon$ as
\begin{eqnarray}
  \mbox{``inner region'':}\hspace*{1cm} \phi& = & \phi^{\rm
    in}_0(\xi_v,{\bf s},t) + \varepsilon \phi^{\rm in}_1(\xi_v,{\bf s},t)
  + \cdots~, \nonumber\\ u & = & u^{\rm in}_0(\xi_v,{\bf s},t) +
  \varepsilon u^{\rm in}_1(\xi_v,{\bf s},t) + \cdots~,\label{inner} \\ 
  \mbox{``outer region'':}\hspace*{1cm} \phi& = & \phi^{\rm
    out}_0({\bf r},t) + \varepsilon \phi^{\rm out}_1({\bf r},t) +
  \cdots~,\nonumber \\ u & = & u^{\rm out}_0({\bf r},t) + \varepsilon
  u^{\rm out}_1({\bf r},t)+\cdots~.\label{outer}
\end{eqnarray}
These ``inner'' and ``outer'' expansions then have to obey matching
conditions \cite{fife2,kupferman} (according to the theory of matched asymptotic
expansions \cite{bender,vandyke}, the outer expansion of the inner
solution has to be equal to the inner expansion of the outer
solution).  We will not discuss these here, but instead
limit ourselves to an analysis of the inner problem\footnote{You may
  easily verify yourself that by substituting (\ref{outer}) into Eqs.
  (\ref{pp1})-(\ref{pp3}) the equation for $\phi^{\rm out}_0$ reduces
  to $g(\phi^{\rm out}_0,u^{\rm out}_0)=0$ which shows that $\phi^{\rm
    out}_0 $ is just ``slaved'' to $u^{\rm out}_0$: to lowest order,
  the order parameter in the bulk (outer) region is the value of
  $\phi^{\rm out}_0$ which minimizes the free energy density $f$ at the local
  temperature $u^{\rm out}_0$. }. On the inner scale, we
have\footnote{You can easily convince yourself of the correctness of
  this result by taking the interface as locally spherical with radius
  of curvature $R$. In spherical coordinates, the radial terms of
  $\nabla^2$ are $\partial^2/\partial r^2 + (2/r) \partial /\partial
  r$, 
  which gives $\varepsilon^2 \nabla^2 \approx \varepsilon^2
  (\partial^2/ \partial r^2 + (2/R) \partial /\partial r) = \partial^2/
  \partial \xi_v + (\varepsilon \kappa ) \partial / \partial \xi_v +
  \cdots$.}
\begin{equation}\label{nablaexp}
  \varepsilon^2 \nabla^2 = {{\partial^2}\over{\partial \xi_v^2}} +
  \varepsilon \kappa {{\partial }\over{\partial \xi_v}} + O(\varepsilon^2)~.
\end{equation}
Furthermore, we shall treat the term $ u$ in $g$ formally as a term of
order $\varepsilon$ and write $v=v_1 \varepsilon+\cdots$ and $u=u_1
\varepsilon +\cdots$ --- this is not so elegant and not necessary
either, but it gets us to the proper answer efficiently. As the
velocity is then also of order $\varepsilon$, this implies that
$\phi_0$ is then the stationary front profile ($\partial \phi_0 /
\partial \xi_v =0$) between two phases in
equilibrium, so that  from (\ref{pp3})  the lowest order equation
becomes
\begin{equation}\label{phi0}
  {{\partial^2 \phi^{\rm in}_0}(\xi_v) \over{\partial \xi_v^2}} + g(\phi^{\rm
    in}_0(\xi_v), 0)=0~.
\end{equation}
The solution of this equation is just the equilibrium profile $\phi_0$
that we introduced in our discussion of the surface tension. Of
course, it is not at all surprising that (\ref{phi0}) emerges in
lowest order, since at $u=0$ the two phases are in equilibrium. Now,
in the next order, we get
\begin{equation}
  \left( {{\partial^2 }\over{\partial \xi_v^2}} + g'(\phi^{\rm
      in}_0) \right) \phi^{\rm in}_1(\xi_v) = -(v_1+\kappa)
    {{\partial \phi^{\rm in}_0(\xi_v) }\over{\partial \xi_v}} - \left.
    {{\partial g(\phi^{\rm in}_0,u)}\over{\partial u}}
  \right|_{u=0} u_1 ~.\label{phi1}
\end{equation}
This equation allows us to solve for $\phi^{\rm in}_1$ in principle.
But even without doing so explicitly, we can get the most important
information out of it. The operator  between parentheses on the left is
nothing but the linear operator we already encountered before: the
Schr\"odinger operator in our discussion of stability. We then saw
that this operator has a mode with eigenvalue zero, the translation
mode $\d \phi_0 /d\xi_v$. Moreover, since the operator is hermitian, it
is also a left eigenmode with eigenvalue zero of this operator. This
implies that for the equation to be solvable, the right hand side has
to be orthogonal to the left zero mode $ \d \phi_0 /d\xi_v$. This
conditions leads to a so-called solvability condition. Upon
multiplying Eq. (\ref{phi1}) by $ \d \phi_0 /d\xi_v$ and integrating,
we can write this condition as an expression for the normal interface
velocity $v_n$ to lowest order in $\varepsilon$,
\begin{equation}\label{solvability}
  v_n = - \kappa - {{\int d\xi_v {{d \phi_0}\over{d\xi_v}} {{ \partial
          g(\phi_0,u)}\over{\partial  u}} ~ u_1}\over{ \int d\xi_v
      \left( {{d \phi_0}\over{d\xi_v}}\right)^2}}~.
\end{equation}
Here, we used the fact that $\phi^{\rm in}_0 = \phi_0$. Moreover, in
the integration on the right hand side, we can take $u$ constant,
since the temperature does not vary to lowest order in the interfacial
region (its derivatives do --- see  \cite{kupferman} for more details).

The above expression is our central result.  The fact that the
prefactor of the curvature term on the right is unity comes from the
fact that the curvature enters according to the expansion
(\ref{nablaexp}) of the diffusion term $\nabla^2$ in precisely the
same way as the velocity term that arises from the transformation to
the co-moving curvilinear from $\xi_v$. When $u_1=0$, i.e., when we
consider an interface between two equilibrium phases, it expresses the
tendency of the interfaces to straighten out. This effect  drives
coarsening  \cite{gunton,bray,langer2,allen}, and the motion is
sometimes refered to as motion by mean curvature. The second term
gives the 
driving term when the interface temperature $u$ is not equal to the
equilibrium temperature. The structure of this term is also quite
transparent. In the denominator, we recognize the surface tension
(\ref{sigma}), and as we already discussed, the inverse of the surface
tension plays the role of an interface mobility in the context of the
type of models we consider. In the numerator we can write $g$ in terms
of $-\partial f(\phi_0,u)/\partial \phi_0$ and then do the integral in
the same way as before in deriving (\ref{vlinear}); we then simply get
\begin{equation}
  v_n= - \kappa - {{1}\over{\sigma}} \left. {{d \Delta f}\over{d u}}
\right|_{u=0} u~.
\end{equation}
where now $\Delta f$ is the difference in free energy densities at
opposite sides of the interface. Clearly, the second term is exactly
what we could have guessed on the basis of what we already knew
before, and together with the curvature term it has exactly the same
type of structure as the boundary condition (\ref{dendrbc2}) that we
introduced in our first discussion of solidification. The
complications that are necessary to model anisotropic kinetics and
surface tension with a phase-field model are significant  \cite{karma},
but conceptually the analysis is essentially the same.

By taking big steps, we have not done justice to the systematics of
the analysis, and there is much more to say about it. If you want to
know more, you will find  entries to the literature in
 \cite{fife,karma,fife2,kupferman}. However, the point we want to bring
to the foreground, following \cite{ebert2} is that in all such
approaches, a hidden assumption 
is made in writing the inner expansion as $\phi^{\rm in}=\phi^{\rm
  in}_0(\rho/ \varepsilon, {\bf s},t)+\cdots $ in (\ref{inner}). In
doing so, we basically already {\em assume} that on the slow time
scale $t$, the profile responds instantaneously to variations in the
outer field $u$. This is why on the inner scale, the changes in the
profile (like $\phi^{\rm in}_1$) are given by {\em ordinary
  differential equations} with coefficients which may vary on the
outer slow time scale. As it happens, this is actually justified for
these type of problems. For, we have seen that the relaxation of a
profile goes exponentially fast, as the spectrum of temporal
eigenvalues $E$ has a finite gap $\Delta E$. In the present case,
where the time scale $\tau$ in the order parameter equation scales as
$\varepsilon$, this means that the relaxation of the front profile
goes as $e^{-\Delta Et/\varepsilon}$. This shows that as $\varepsilon
\rightarrow 0$, the adiabatic assumption implicit in the above
analysis {\em is right}, as the relaxation on the inner scale
completely decouples from the slow scale variation of the outer
fields. In other words: we have left out exponentially small terms as
$\varepsilon \rightarrow 0$, but that is something that almost always
happens when we an asymptotic expansion!  As we shall see now, the
adiabatic approximation can not be made for fronts moving into an
unstable state, such as streamers\footnote{At the summerschool, Roger
  Folch Manzanares nicely illustrated  to me  how one can
  go wrong with an effective interface approximation if one does not
  think about the stability of the equations 
  on the inner scale: in a first naieve attempt to formulate phase
  field 
  equations for the viscous finger problem, he had explored equations
  which did reduce to the standard viscous finger equations if one
  blindly followed the standard recipe for analyzing the
  $\varepsilon 
  \rightarrow 0$ limit. However, the coupling of the phase field with
  the outer pressure-like field was such that the equations were
  completely unstable on the inner scale for small $\varepsilon$. So do watch out!}.

\section{Some elements of front propagation into unstable states ---
  relaxation and the effective interface
  approximation}\label{theme3}

We now briefly touch on a few elements of fronts propagating into
unstable states. In view of the length restrictions on the
contribution to the proceedings of the school, we only highlight some
recent results obtained in collaboration with Ebert  \cite{ebert2},
which show that a large class of fronts propagating into an unstable
state show universal power law relaxation and that this makes the
mapping of such fronts onto an effective interface model questionable.

Our own motivation comes from our attempt to understand the streamer
problem, but examples of fronts propagating into an unstable state
arise in various fields of physics: they are important in many
convective instabilities in fluid dynamics such as the onset of von
Karman vortex generation  \cite{provansal}, in Taylor  \cite{ahlers} and
Rayleigh-B\'enard  \cite{fineberg} convection, they play a role in
spinodal decomposition near a wall  \cite{ball}, the pearling
instability of laser-tweezed membranes  \cite{goldstein}, the formation
of kinetic, transient microstructures in structural phase transitions
 \cite{salje}, the propagation of a superconducting front into an
unstable normal metal  \cite{dorsey}, or in error propagation in
extended chaotic systems  \cite{torcini}.  The experimental relevance
of the understanding of the relaxation of such fronts is illustrated
on propagating Taylor vortex fronts. Here the measured velocities were
about 40\% lower than predicted theoretically \cite{ahlers}, and only
later numerical simulations  \cite{luecke} showed that this was due to
slow transients.

When one of the states is unstable, even a small perturbation around
this state will grow out and spread; therefore, the properties of
fronts that propagate into an unstable state depend on the initial
conditions.  If the initial profile is steep enough, arising, e.g.,
through local initial perturbations, it is known that the propagating
front in practice always relaxes to a unique profile and
velocity \cite{aw,benjacob,vs1,vs2,oono}. Depending on the
nonlinearities, one generally can distinguish two regimes: as a rule,
fronts whose propagation is driven (``pushed'') by the nonlinearities,
resemble very much the fronts which propagate into a metastable state
and which we have discussed extensively in section \ref{theme2} (e.g.,
their relaxation is also exponential in time). We will therefore not
consider this regime, which is often refered to as ``pushed''
 \cite{stokes,oono} or ``nonlinear marginal stability''  \cite{vs2} any
further. If, on the other hand, nonlinearities mainly cause
saturation, fronts propagate with a velocity determined by
linearization about the unstable state, as if they are ``pulled'' by
the linear stability (``pulled'' \cite{stokes,oono} or ``linear
marginal stability'' \cite{vs1,vs2} regime).

\begin{figure}
\epsfxsize=13.6cm
\epsffile{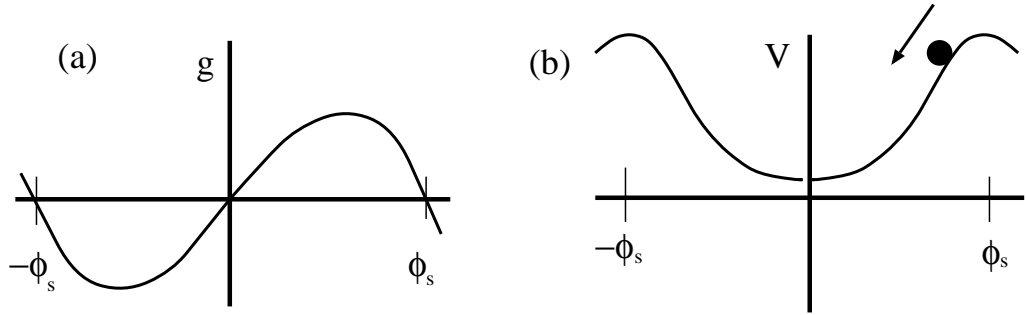}
\caption{\footnotesize\label{gunst}The functions $g$ and $V$ in the case of front
  propagation into unstable states. Compare Fig. \ref{gfv}, where the
  functions are drawn for the case of a front between a stable and a
  metastable state.}
\end{figure}

Almost all important differences between ``pulled'' or ``linear 
marginal stability'' fronts propagating into an unstable state and
those propagating into a  metastable state trace back to the fact that in the latter
case there typically is a discrete set of front solutions, only one of
which is stable as was illustrated in Fig. \ref{vis}, while in the 
former case there generally is a family of moving front solutions
 \cite{aw,benjacob,vs1,vs2}. To illustrate this, we again turn to Eq. 
(\ref{op1}), but now take $g(\phi)$ of the form sketched in Fig.
\ref{gunst}{\em (a)}. In this case, $g'(0)>0$, so the state $\phi=0$
is unstable. If we again consider fronts propagating into this state,
the potential $V$ corresponding to this function $g$ is the one shown
in Fig. \ref{gunst}{\em (b)}. Now, the question of the existence of a
uniformly translating profile $\phi_v(x-vt)=\phi(\xi_v)$ translates into
the question ``is there a solution in the particle on the hill analogy
in which the particle starts at time $\xi_v=-\infty$ at the top, and
comes to the bottom as $\xi_v\rightarrow \infty$?''.  Obviously, there is
such a solution for {\em any} positive value of the ``friction
coefficient'' 
$v$: {\em there is a continuous family of uniformly translating front
  solutions}.

It is useful to consider the relation between the velocity $v$ which
labels the front solutions, and the asymptotic decay rate $\Lambda$:
if we linearize Eq. (\ref{op1}) around the state $\phi=0$ and write
$\phi_v \sim e^{-\Lambda \xi_v}$, then we get $v \Lambda = \Lambda^2
+g'(0)$, so
\begin{equation}\label{lambda}
  \Lambda_\pm = {{v}\over{2}} \pm \sqrt{{{v^2}\over{4}} -g'(0)}~,
  \hspace*{1.5cm} g'(0)>0~.
\end{equation}
For $v>2g'(0)$, the roots are real, and $\Lambda_- <\Lambda_+$. For
$v<2g'(0)$, the roots are complex, meaning that the front solutions
decay to zero as $\phi_v \sim \cos(\Im \Lambda_\pm \xi_v) e^{-\Re
  \Lambda_\pm \xi_v}$. Clearly, the velocity $v^*=2g'(0)$ is a special
value, as the two roots coincide there
$\Lambda_-=\Lambda_+=\Lambda^*$. It is a well-known result that in
such a degenerate case, the front profile does not decay as a single
exponential, but that instead in this case
\begin{equation}\label{phi*}
  \phi_{v^*}(\xi_v) \sim (\xi_v + \mbox{const.} ) e^{-\Lambda^*\xi_v}~.
\end{equation}
so that the dominant behavior for large $\xi_v$ is the $\xi_v
e^{-\Lambda^* \xi_v}$ term.

The special status of the value $v^*$ also becomes clear when we look
at the stability analysis of the fronts $\phi_v(\xi_v)$. If we retrace
the stability analysis of section \ref{theme2}, then in this case the
potential $U(\xi_v)$ in the Schr\"odinger type equation for the spectrum
has an asymptotic value $(v^*)^2/4-g'(0)=0$. Hence, according to our
arguments the continous spectrum associated with quantum mechanically
allowable eigenfunctions\footnote{At this point, the warning of
  footnote \ref{footnotewarning} on page \pageref{footnotewarning}
  becomes important: for fronts propagating into an unstable state,
  there are important eigenfunctions of the stability operator which
  are not in the class of eigenfunctions that are allowed in quantum
  mechanics, as they diverge as $\xi_v \rightarrow \pm \infty$. These
  are especially important when studying the stability of front
  solutions with velocity $v>v^*$, as these are the type of solutions
  whose eigenvalue continues all the way down to zero. As a result,
  the stability spectrum is always gapless. See  \cite{ebert2} for
  further details.} $\psi$ comes all the way down to zero, i.e., there
is no gap. This already gives a hint that there will be
non-exponential relaxation.

In the derivation of effective interface equations, we encountered
solvability conditions which involved integrals of the form $\int dx
(d\phi_0/dx)^2 $ --- see Eq. (\ref{solvability}).  In the present
case, the front velocity is always nonzero, and as a result the
stability operator is non-hermitian \cite{ebert2}. If one tries to
derive effective 
interface equations for such fronts using the same type of approach as
discussed at the end of section \ref{theme2}, one needs the zero mode
of the adjoint operator of the problem with $v\neq 0$ in the
corresponding 
solvability condition. 
Because of the non-hermitian nature of this operator for $v\neq 0$,
this   zero mode 
turns out to be $e^{v\xi_v} (\d\phi_v/d\xi_v)$, and one encounters
integrals of the type $\int d\xi_v e^{v\xi_v} (d\phi_v/d\xi_v)^2$
(note that for $v=0$, the zero mode of the adjoint operator reduces to
the one we used before, 
$\partial \phi_0 / \partial \xi_v$). As  
$\xi_v\rightarrow \infty$, the integrand behaves as\footnote{The
  factor ${\sqrt{v^2/4-g'(0)}}$ in the exponential is zero at
    $v^*$. At $v^*$, the integrals still diverge, but only as
    a power law \cite{ebert2}.} 
$e^{(v-2\Lambda_-)\xi_v}= e^{\sqrt{v^2/4-g'(0)}\xi_v}$. As a result, the
integrals that arise if one naievely applies the standard analysis do
not converge.  Although there have been some suggestions  \cite{chen}
that one might regularize such integrals by introducing a cutoff which
is taken to infinity at the end of the calculations, such fixes do not
appear to work here and obscure the connection of this problem with
the slow relaxation discussed below.

We have not yet discussed the origin of the result that ``pulled'' fronts which emerge from
sufficiently localized initial conditions move with a speed  
$v^*$ determined by the linear behavior of the dynamical equation [in
our case, the fact that $v^*$ is determined solely by $g'(0)$]. The
origin   
lies in 
the fact that any perturbation about the unstable state grows out and
spreads by itself. This leads to a natural spreading speed of linear
perturbations, and $v^*$ is nothing but this speed itself  
 \cite{vs2,landau}. If nonlinearities mainly suppress further  growth, then
indeed the dynamically relevant front is ``pulled''  \cite{stokes} by
the leading edge whose dynamics is governed by the linearized
equation.  Ebert and I have recently found that one can build on this
idea to analyze the relaxation of front profiles towards $\phi_{v^*}$
 \cite{ebert2}. The main idea can be illustrated within the context of
the dynamical equation (\ref{op1}) as follows. Let us use the freedom
of choosing appropriate space and time scales to take $g'(0)=1$. As
the discussion following Eq. (\ref{lambda}) shows, $v^*=2$ and
$\Lambda^*=1$ in this case, and the linearized dynamical equation 
reads
\begin{equation}
  {{\partial \phi(x,t) }\over{\partial t}}={{\partial^2
      \phi(x,t)}\over{\partial x^2}} + \phi(x,t)~.
\end{equation}
We now write the equation in the moving frame $\xi_v=x-v^*t$ moving with
velocity $v^*=2$, and make the transformation
$\phi(\xi_v,t)=e^{-\xi_v}\psi(\xi_v,t)$. This is essentially the same type of
transformation that we did before in (\ref{psi}) when we performed the
stability analysis of moving front solutions. With these
transformations, $\psi$ simply obeys the diffusion equation 
\begin{equation}
  {{\partial \psi (\xi_v,t) }\over{\partial t}}={{\partial^2
      \psi(\xi_v,t)}\over{\partial x^2}}~.
\end{equation}
As is well known, in many diffusion type problems the long time 
asymptotics is governed by the fundamental similarity solution or one
of its derivatives, like 
\begin{equation}
  \psi^{\rm sym}_1= {{1}\over{ {t}^{1/2}}} e^{-\xi_v^2/4t}
  ~,\hspace*{0.5cm}\mbox{or}\hspace*{0.5cm} \psi^{\rm sym}_2=
  -{{\partial \psi^{\rm sym}_1}\over{\partial \xi_v}}= {{\xi_v}\over{2\ 
      t^{3/2}}} e^{-\xi_v^2/4t}~,
\end{equation}
so it is not unreasonable to expect that one of these similarity 
solutions also governs the long time asymptotics in the leading edge
here. If so, the corresponding function $\phi(\xi_v,t)$ should approach
the dominant $\xi_ve^{-\xi_v}$ term of (\ref{phi*}) for
large times.  As $\psi=e^{\Lambda^*\xi_v} \phi$, this means that the
spatial dependence of the similarity solution $\psi^{\rm sym}$ that we
are looking for should go as $\xi_v$ for $\xi_v^2 \ll t$. Clearly, the
appropriate one is $\psi^{\rm sym}_2$. Hence, this simple argument
suggests that in the frame moving with velocity $v^*=2$, the dominant
long time dynamics in the leading edge is 
\begin{equation}
  \phi \sim {{\xi_v}\over{t^{3/2}}} e^{-\xi_v -\xi_v^2/4t} = e^{-\xi_v-3/2 \ln
    t + \ln \xi_v -\xi_v^2/4t}~.\label{asymp}
\end{equation}
If we now track the position $\xi_h(t)$ of the point where
$\phi(\xi_v,t)=h$, we get to dominant order from the requirement that the
exponent in the above expression remains constant 
\begin{equation}
  \xi_h = -{{3}\over{2}} \ln t + \cdots
  \hspace*{1cm}\Longleftrightarrow \hspace*{1cm} \dot{\xi}_h =
  -{{3}\over{2t}} + \cdots
\end{equation}
As $\dot{\xi}_h$ is the velocity of the point where $\phi=h$, we see
that in the leading edge of the profile the velocity relaxes towards
$v^*$ as $-3/(2t)$. This is precisely what was found by Bramson
 \cite{bramson} from a rigorous analysis. Although the above argument
is rather handwaving, we have recently found  \cite{ebert2} that it can
be made into a systematic asymptotic analysis which applies not just 
to the second order dynamical equation (\ref{op1}), but also to higher
order partial differential equations which admit uniformly translating
front solutions. The surprising finding is that not just the leading
order $\sim 1/t$ relaxation term in the velocity is universal, but
also the first subdominant $\sim 1/t^{3/2}$ term, which can not be
obtained from the sove argument: independent of the
``height'' $h$ whose position we track, we find that the velocity
$v_h(t)=v^*+\dot{\xi}_h$ relaxes to $v^*$ as 
\begin{equation} 
  v_h=v^* - {{3}\over{2\Lambda^* t}} \left(1-\frac{\sqrt{\pi}}{\Lambda^*
    \sqrt{Dt}} \right) + O\left(\frac{1}{t^2}\right)~,
\end{equation}
where for the order parameter equation (\ref{op1}) with $g'(0)=1$,
$v^*=2 $, $\Lambda^*=1$ and $D=1$. In the more general case, $D$ is a
coefficient which plays the role of a  diffusion coefficient, and
which can be determined explicitly from the dispersion relation of the
linearized equation.
Moreover, also the shape of the profile relaxes 
with the same slow power laws in a universal way which is related to
the existence of a family of front solutions.  We refer to
 \cite{ebert2} for details. 

The above $1/t$ power law relaxation is clearly too slow to make an
effective interface approximation with boundary conditions which are
local in space and {\em time} possible for ``pulled'' fronts whose
propagation into an unstable states originates in diffusive spreading
and growth. To see this, consider, e.g., a spherically symmetric front
in Eq. (\ref{op1}) in three dimensions which grows out from some
localized region around the origin. For long times the front region is
thin in comparison with the distance $r_f$ from the origin and the
curvature of the front is small and of order $2/r_f\approx 2/(2v^*t)$.
Thus, the curvature is of the same order as the dominant relaxation
term of the front, and one can not simply express the instantaneous
front velocity in terms of $v^*$ plus some kind of curvature
correction, as we saw one can do for fronts between a stable and a
metastable state. Some preliminary numerical investigations have
confirmed this. Whether some other interfacial description with memory
type of terms can be developed, or whether there are other unexpected
consequences of this slow relaxation, is at present an open 
question.

\section*{Acknowledgement}
Much of my thinking on the issues discussed in these lecture notes has
been shaped by interactions and collaborations with Christiane Caroli
and Ute Ebert. I wish to thank both of them. In addition, I want to
thank Ute Ebert and Ramses van Zon for extensive comments on an
earlier version of the manuscript, and Lucas du Croo de Jongh for
teaching me how to make computer drawings.

\end{document}